\documentclass[twocolumn]{aastex62}
\usepackage{ulem}

\graphicspath{{./}{figures/}}

\received{January 1, 2018}
\revised{January 7, 2018}
\accepted{\today}

\submitjournal{ApJ}

\shorttitle{Constraining the gap size of HD\,100546}
\shortauthors{Jamialahmadi et al.}

\begin{document}

\title{Constraining the gap size in the disk around HD\,100546 in the mid-infrared}

\correspondingauthor{Narges Jamialahmadi}
\email{nari.jami@gmail.com, jami@oca.eu}

\author{Narges Jamialahmadi}
\affil{School of Astronomy, Institute for Research in Fundamental Sciences (IPM), P.O. Box 19395-5746, Tehran, Iran}

\author{Thorsten Ratzka}
\affiliation{Institute for Physics / IGAM, NAWI Graz, Karl-Franzens-Universit{\"a}t, 
Universit{\"a}tsplatz 5/II, 8010 Graz, Austria}

\author{Olja Pani\'c}\footnote{Royal Society Dorothy Hodgkin Fellow}
\affiliation{School of Physics and Astronomy, University of Leeds, Woodhouse Lance, LS29JT Leeds, United Kingdom}

\author{Hassan Fathivavsari}
\affiliation{School of Astronomy, Institute for Research in Fundamental Sciences (IPM), P.O. Box 19395-5746, Tehran, Iran}n

\author{ Roy van Boekel}
\affiliation{Max Planck-Institut f{\"u}r Astronomie, K{\"o}nigstuhl 17, D-69117 Heidelberg, Germany}

\author{Sebastien Flement}
\affiliation{Laboratoire J.-L. Lagrange, UMR 7293, University of Nice Sophia-Antipolis, CNRS, \\  Observatoire de la Cote D‚Äö√Ñ√¥Azur, Boulevard de l‚Äö√Ñ√¥Observatoire - CS 34229 - F 06304 NICE Cedex 4, France}

\author{ Thomas Henning}
\affiliation{Max Planck-Institut f{\"u}r Astronomie, K{\"o}nigstuhl 17, D-69117 Heidelberg, Germany}

\author{ Walter Jaffe}
\affiliation{Sterrewacht Leiden, Niels Bohrweg 2, 2333 CA Leiden, The Netherlands}

\author{Gijs D. Mulders}
\affiliation{Lunar and Planetary Laboratory, The University of Arizona, Tucson, AZ 85721, USA Earths \\ in Other Solar Systems Team 
, NASA Nexus for Exoplanet System Science}

\begin{abstract}
We refine the gap size measurements of the disk surrounding the Herbig Ae star HD\,100546 in the N band. Our new mid-infrared interferometric (MIDI) data have been taken with the UT baselines and span the full range of orientations. The correlated fluxes show a wavy pattern in which the minima separation links to a geometrical structure in the disk. We fit each correlated flux measurement with a spline function, deriving the corresponding spatial scale, while assuming that the pattern arises interferometrically due to the bright emission from the inner disk and the opposing sides of the wall of the outer disk. We then fit an ellipse to the derived separations at their corresponding position angles, thereby using the observations to constrain the disk inclination to $i=$47\,$\pm$\,1$^{\circ }$ and the disk position angle to $PA=$135.0\,$\pm$\,2.5$^{\circ}$ East of North, both of which are consistent with the estimated values in previous studies. We also derive the radius of the ellipse to 15.7\,$\pm$\,0.8\,au. To confirm that the minima separations translate to a geometrical structure in the disk, we model the disk of HD\,100546 using a semi-analytical approach taking into account the temperature and optical depth gradients. Using this model, we simultaneously reproduce the level and the minima of the correlated fluxes and constrain the gap size of the disk for each observation. The values obtained for the projected gap size in different orientations are consistent with the separation found by the geometrical model.
\end{abstract}


\section{Introduction} \label{sec:intro}
 
 The protoplanetary disks surrounding Young Stellar
Objects (YSOs) evolve by various physical
mechanisms such as photoevaporation \citep[]{1994ApJ...428..654H,2006MNRAS.369..229A,2011ARA&A..49..195A}, accretion/ejection
\citep[]{1974MNRAS.168..603L,2011ARA&A..49..195A,2014prpl.conf..475A} and dust growth and planet formation \citep[]{2007MNRAS.377.1324C,2014prpl.conf..497E,2014prpl.conf..339T}. These processes can occour at the same time
with timescales of up to a few million years and
can cause substantial evolution in the disks such as
clearing disk cavities. In this case, the disks are named the transitional disks.

The first transitional disk was recognized as a
subclass of protoplanetary disks by \citet{1989AJ.....97.1451S} and \citet{1989AJ.....98.1409S} from near-infrared
(NIR) ground-based photometry and IRAS
(MIR) photometry. Detailed studies of these objects
became feasible as progressively more sophisticated
instruments, the Spitzer Space Telescops \citep{2004ApJS..154....1W} identified a new class of disks called
pre-transitional disks around stars LKCa 15 and UX
Tau A \citep{2007ApJ...670L.135E}.
The spectral energy distribution (SED) of such
disks, which provide indirect evidence of gaps have
lack of MIR flux (5--20 $\mu$m ) and a significant
excess at NIR (2--5 $\mu$m) and longer wavelengths
compared to the full protoplanetary disks. The
specified shapes of these SEDs demonstrates that
pre-transitional disks are comprised of an inner
disk and an outer disk separated by a gap. One of
the proposed formation scenarios of the gapped
disks are the disk-planet interaction while the
geometrical structure of such gaps are still under
debate.

 HD\,100546 \citep[110\,$\pm$\,0.6\,pc from the second Gaia Data Release,][]{2016A&A...595A...1G,2018arXiv180409365G} hosts a widely studied pre-transitional disk, which was first identified by \citet{2003A&A...401..577B} modeling the spectral energy distribution (SED). They concluded that HD\,100546 has a small inner disk of a few au and a more massive outer disk beyond $\sim$\,10\,au. They speculated about the presence of a planetary companion in the gap. The gas and dust-rich disk of HD\,100546 is one of the best laboratories to study forming planetary systems. Direct and indirect observations reveal that the disk hosts two embedded companion candidates \citep[e.g.][]{2013ApJ...766L...1Q, 2015ApJ...807...64Q,2003A&A...401..577B,2014ApJ...791..136B,2015ApJ...814L..27C}.


The geometrical structures of the disk around HD\,100546 were widely studied in multi-wavelength observations. The results of the various studies are listed in Table\,1. Please note that in the literature the distance of 97\,pc is commonly used. In this work, we use the updated distance of 110\,pc, measured by GAIA. 

As a complementary study, in this paper, we present our mid-infrared interferometric data, consisting of 15 independent two-element baseline measurements obtained with the Very Large Telescope Interferometer (VLTI) and its instrument MIDI \citep{2003Msngr.112...13L}. With a large span of the projected position angles, these data provide a unique measurements of the gap geometry in the mid-infrared in unprecedented detail. It is unique since we have multi-epoch N-band data, which has not been done before. These data give us more information about the geometry of the disk than previous N-band observations on HD 100546. Constraining the precise size of the gap of HD\,100546 at various wavelengths is important to learn about the processes at work that form this gap. Moreover, HD\,100546 exhibits multiple signposts of planets and will be a prime target for all future instruments.

\section{Observations and data reduction} \label{sec:style}

\begin{figure}
\centering
\begin{tabular}{c}
\includegraphics[clip=,width=0.8\hsize]{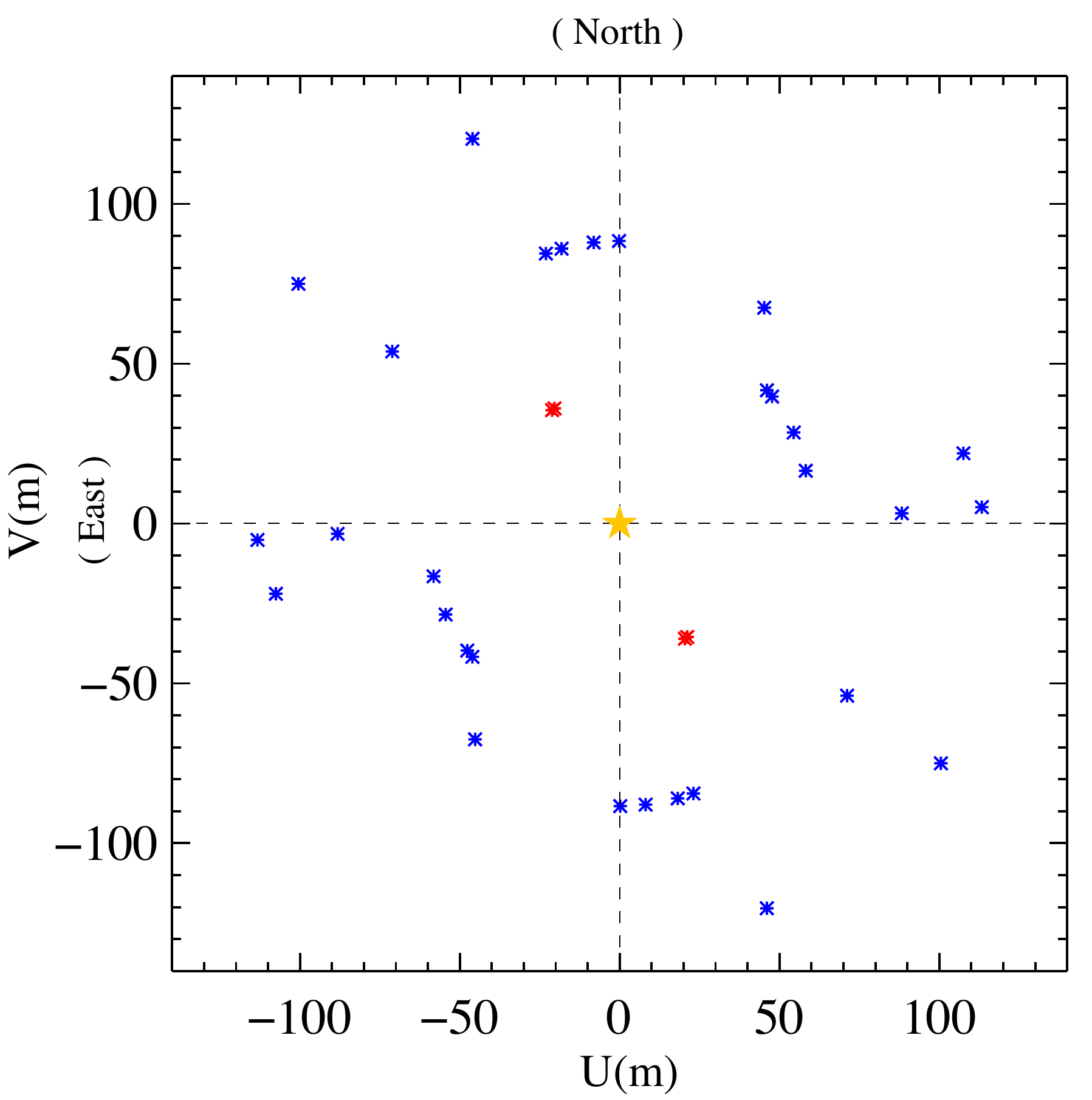} 
\end{tabular}

\caption{The uv-coverage of our UT data from 2012/2013 (blue) and 2004/2005 \citep[red;][]{2014A&A...562A.101P,2003Msngr.112...13L}.}
 \label{1}
\end{figure}

HD\,100546 was interferometrically observed in the N band ($8 - 13\,\mu$m) in 2012 and 2013 using the MIDI/VLTI (ESO GTO program 090.C--0931(A)-(E), PI: O. Pani\'c). In this paper, we focus on our observations carried out with the 8\,m Unit Telescopes (UTs). We performed the observations  with the prism as dispersive element (R\,$\sim$\,30) and in the HIGH-SENS mode, i.e., the total fluxes and the correlated fluxes were measured consecutively. Table\,\ref{table:2} gives the log of observations and Fig.\,\ref{1} presents the corresponding uv-coverage.
 For the calibration of the correlated fluxes and the determination of the transfer function, calibrator stars were observed within about half an hour before or after the target in the same region of the sky. We used the calibrator stars before and after the target and then averaged the results. In the cases for which the transfer function of one of the calibrators was peculiar we used only one calibrator (Table\,\ref{table:3}).

\begin{table*}
\centering
\label{table:1} 
\caption{A comparison between the geometrical parameters of the disk of HD\,100546 from various studies. Listed are the inclination, i, the position angle,PA, and the cavity size of the disk. The absolute sizes of the cavity have been derived for a common distance of 110\,pc (for further details see text).}
\begin{tabular} {p{3.2cm} p{2.5cm}p{2.5cm}p{1.6cm}p{1.8cm}p{2cm}p{2.2cm}}
\hline\hline
 \textbf{Reference} &\textbf{Instrument}& \textbf{wavelength range} & \textbf{i$(^{\circ})$} & \textbf{ PA$(^{\circ})$} &  \textbf{physical Cavity size\,(au)} &  \textbf{corresponding Cavity size at 110 pc (au)} \\ [0.5ex] 
\hline
 \citet{2011ApJ...738...23Q} & VLT/NACO &H and K band &47.0\,$\pm$\,2.7 & 138.0\,$\pm$\,3.9& $\sim$\,15.0 & $\sim$\,16.9  \\ 
 \citet{2014ApJ...790...56A} &VLT/NACO &L band& --- & 138\,$\pm$\,3& 14.0\,$\pm$\,2.0 & 15.8\,$\pm$\,2.3   \\ 
 Garufi et al. (2016) &VLT/SPHERE&Visible&--- & ---& 11.0\,$\pm$\,1.0\ & 12.4\,$\pm$\,1.1\\ 
\citet{2014ApJ...791..136B} & VLT/CRIRES &NIR spectroscopy&42 &---& 12.9\,$_{-1.3}^{+1.5}$ & 14.6\,$_{-1.5}^{+1.7}$ \\ 
 Mulders et al. (2013) &VLTI/MIDI&MIR &53 & 145& $\sim\,10.0\,to\,$\,25.0 & $\sim\,11.3\,to\,$\,28.2 \\  
 Pani\'c et al. (2014) &VLTI/MIDI&MIR &53\,$\pm$\,8 & 145\,$\pm$\,5& 11.0\,$\pm$\,1.0& 12.4\,$\pm$\,1.2\\  
 Menu et al. (2015) &VLTI/MIDI& MIR&--- &--- & 18.8\,$\pm$\,1.2 & 21.2\,$\pm$\,1.4\\ 
 \citet{2014ApJ...788L..34P} &ALMA& 870\,$\mu$m&41.94\,$\pm$\,0.03 &145.14\,$\pm$\,0.04 & --- & ---\\
 \citet{2015MNRAS.453..414W} &ATCA& 7\,mm&40\,$\pm$\,5 &140\,$\pm$\,5 & 25.0 & 28.2\\
 \citet{2017AJ....153..264F} &MagAO and GPI& optical and NIR&42&145& 15.0 & 15.0\\
 Our work &VLTI/MIDI& MIR&47\,$\pm$\,1&135\,$\pm$\,2.5& 15.7\,$\pm$\,0.8 & 15.7\,$\pm$\,0.8\\

     \hline
\end{tabular}
\end{table*}

To reduce the data, we used v2.0 of MIA+EWS\footnote{\url{http://www.strw.leidenuniv.nl/~nevec/MIDI/}}. For weak correlated fluxes the EWS branch with its shift-and-add algorithm in the complex plane is more accurate. In this paper, MIA with its power spectrum analysis was only used to check the EWS results.

We present some of the calibrated correlated fluxes (black lines) with better quality in Fig.\,\ref{2} and the remaining measurements in Fig.\,\ref{alldata} in the online material. 

\section{Results} \label{sec:floats}
 

\subsection{Geometrical model}

The observed correlated fluxes show a wavy pattern with clear minima, which vary with baseline length and orientation. In interferometric observations a sinusoidal signal in the correlated flux is a sign of binary-like structure in the sky \citep[see][]{2009A&A...502..623R,2006A&A...455.1009C}.
Therefore, the most elementary interpretation of the minima in our data are two separated point sources in the sky, i.e., a binary or a structure similar to a binary that causes the minima. If the wavelength-dependence of the separation is weak, the separation can be directly accessed. The only parameter influencing the location of the minima is the angular separation.  Here the inner disk and the opposite sides of the disk wall can mimic a binary when observed with an interferometer.

For each observation, we constrain the position of two consecutive minima and derive the angular gap size along the baseline direction. Afterwards, we fit an ellipse to all the separations with the aim to derive the position angle, the inclination and the size of the gap. For  the given distance, the angular separation can be converted to au.

The identification of the minima locations requires some efforts since the sampling of the MIDI data is quite low and the minima are not always symmetric. Therefore, we try to fit a spline function to the correlated fluxes in each dataset before identifying its corresponding minima. The spline function enables us to predict the correlated flux also in wavelength ranges affected by noise. 
We found that by different selection points used in our spline fit around each minimum, the minima positions change slightly. We repeat this process 100 times and each time record the minima from the fit. The mean of these 100 minima is taken as the best minimum, their standard deviation provides the uncertainty of this minimum. The minima of the spline fit are then taken as the minima of the corresponding correlated flux. The spatial frequencies at which two consecutive minima occur are then subtracted and converted to au for the given distance.

\begin{figure*}
\centering
\begin{tabular}{c}
\includegraphics[clip=,width=0.3\hsize]{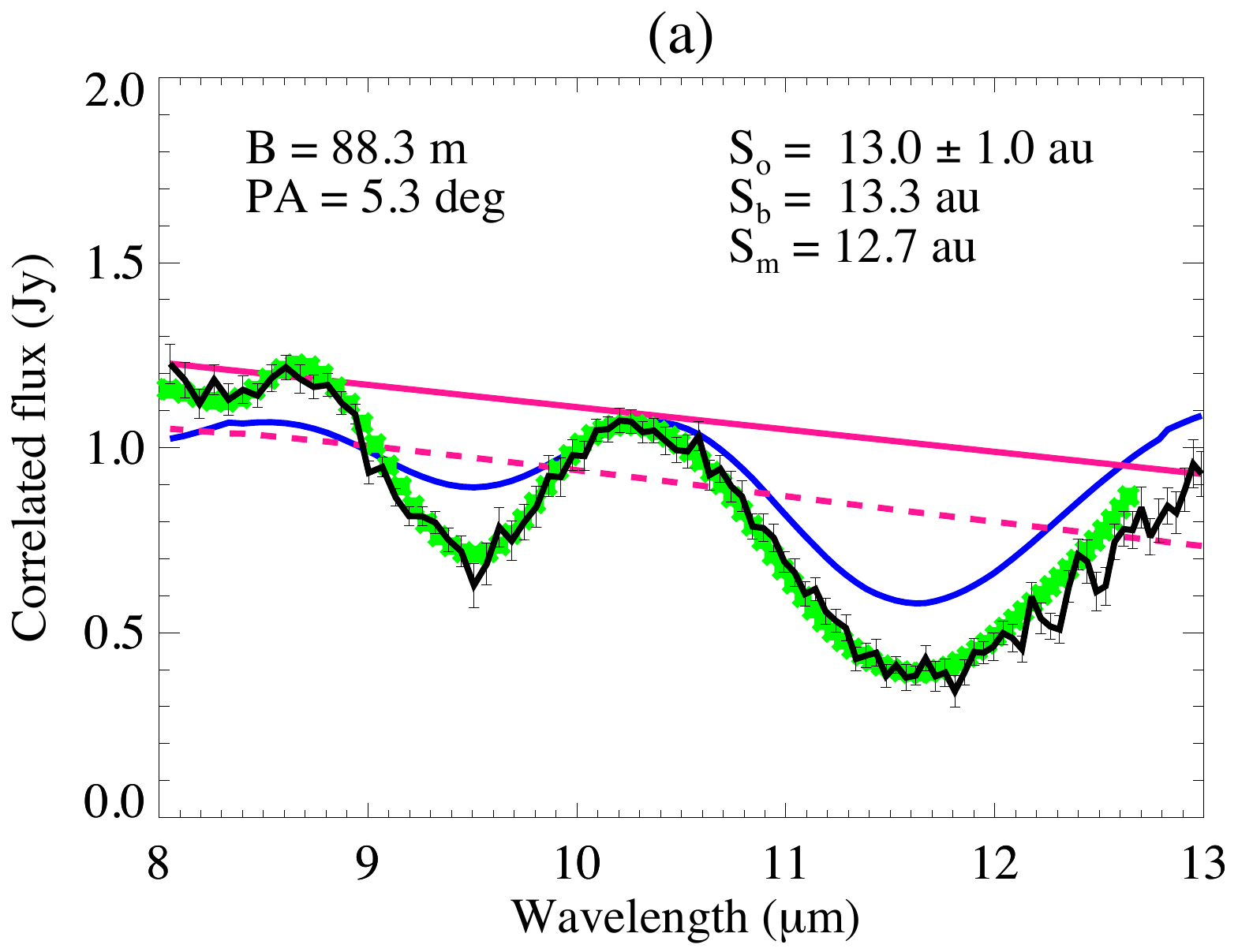} 
\includegraphics[clip=,width=0.3\hsize]{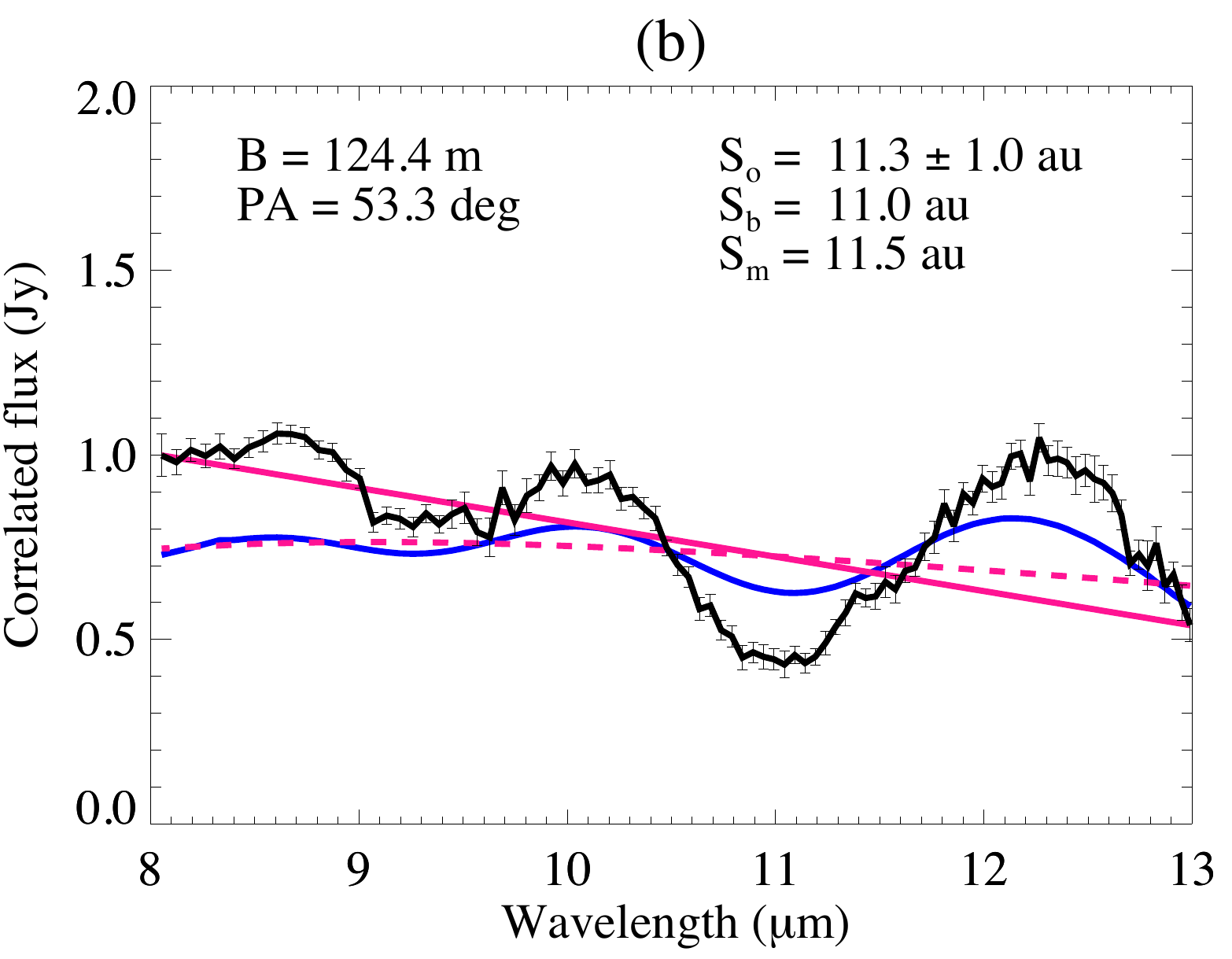} 
\includegraphics[clip=,width=0.3\hsize]{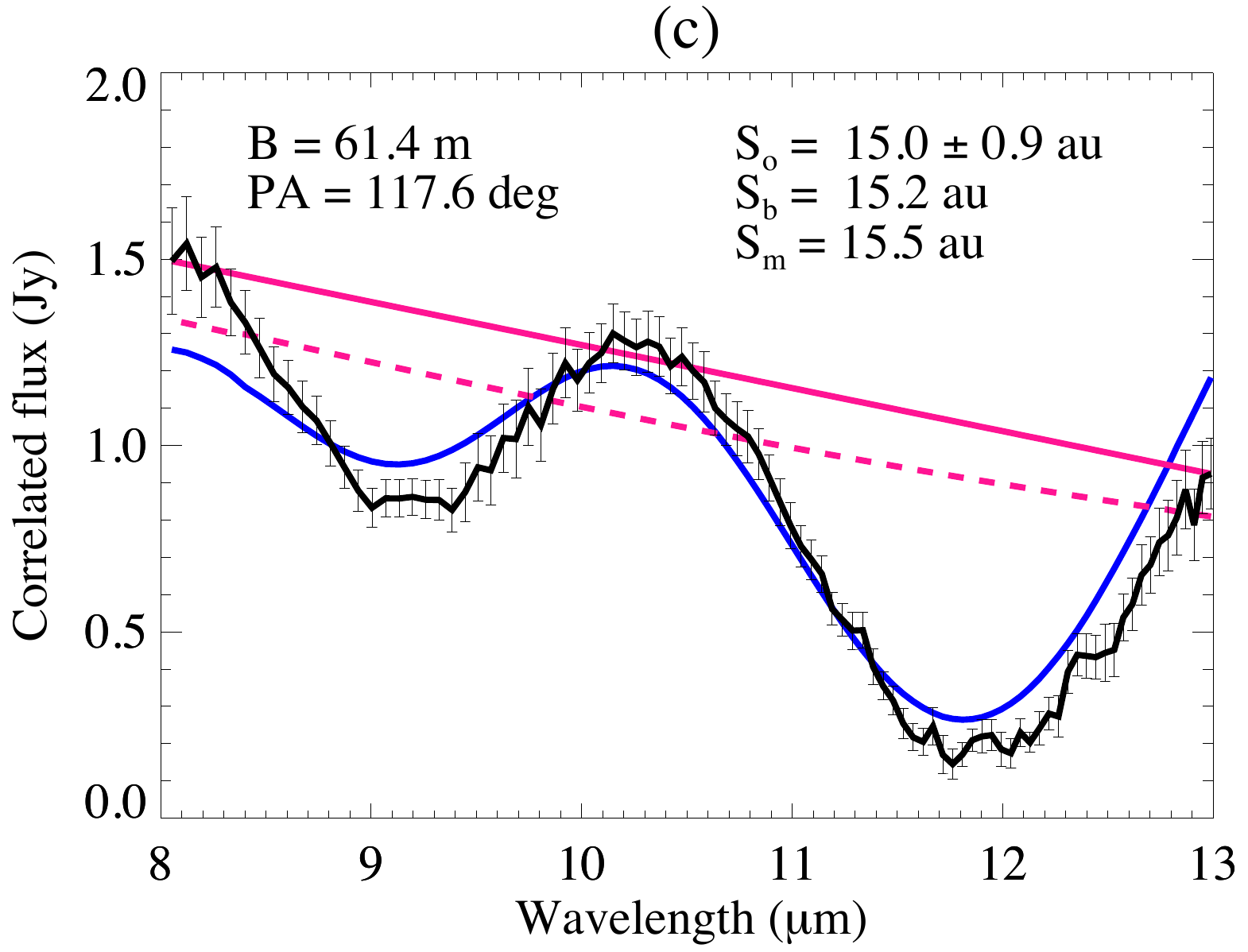}\\
\includegraphics[clip=,width=0.3\hsize]{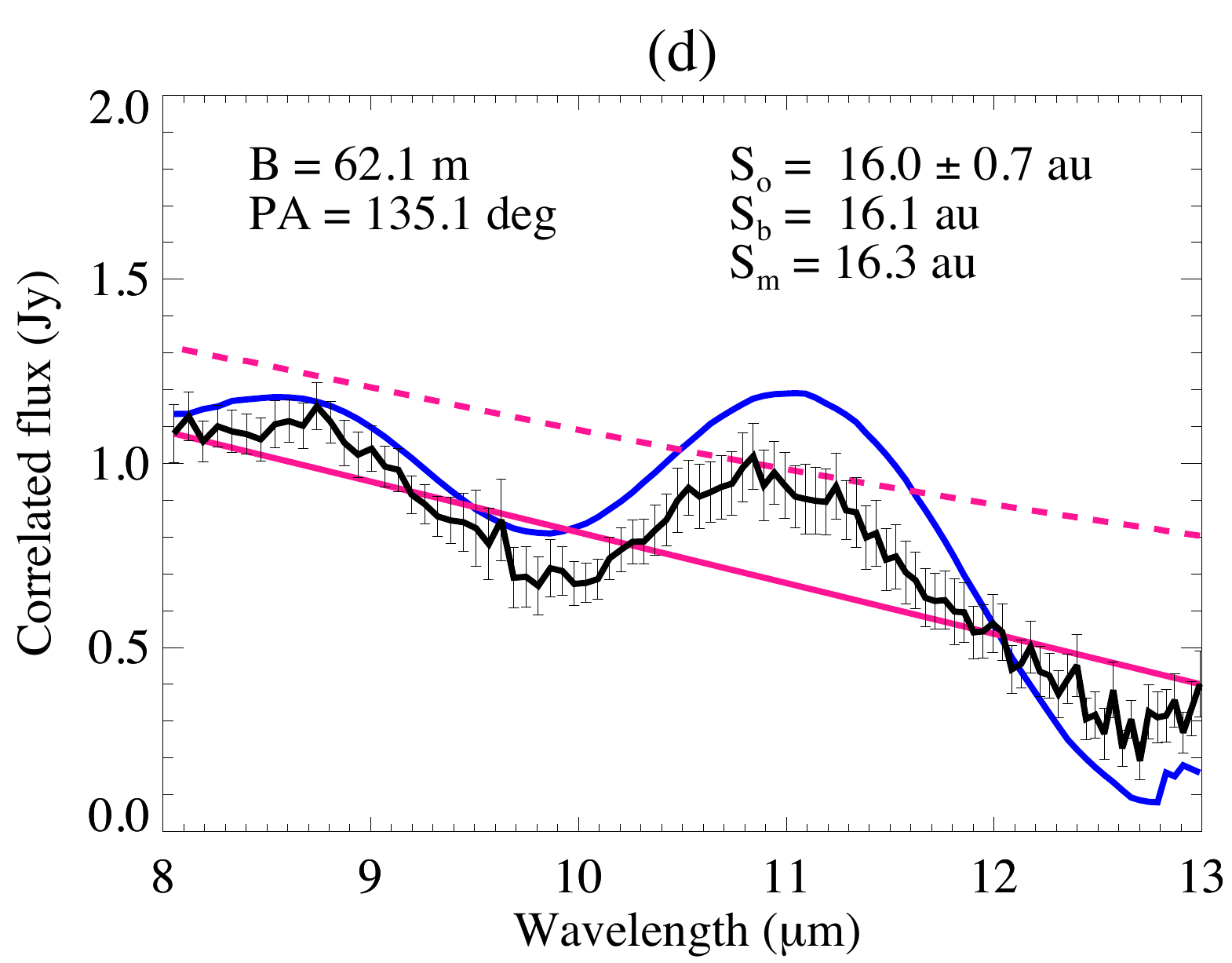} 
\includegraphics[clip=,width=0.29\hsize]{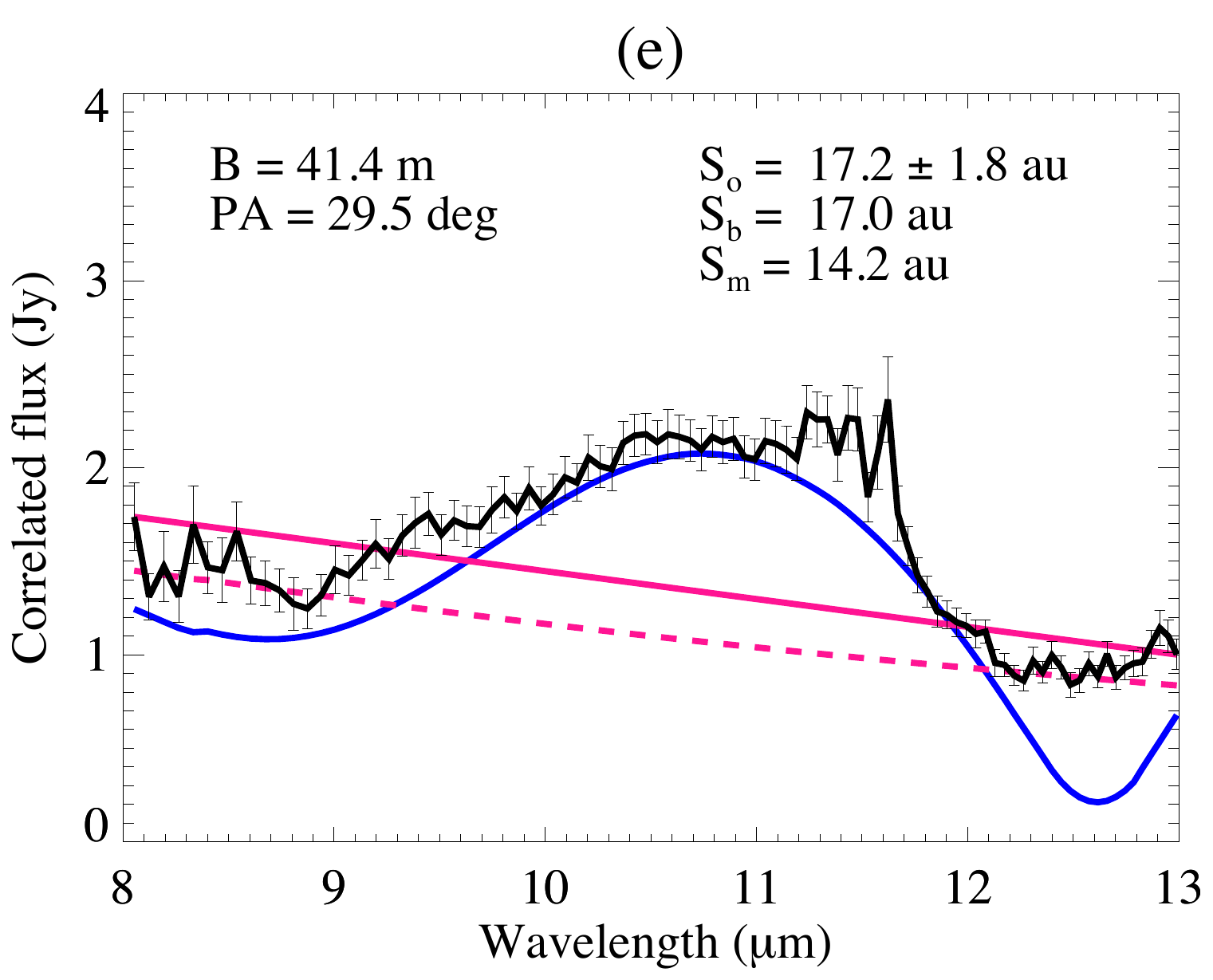}
\includegraphics[clip=,width=0.29\hsize]{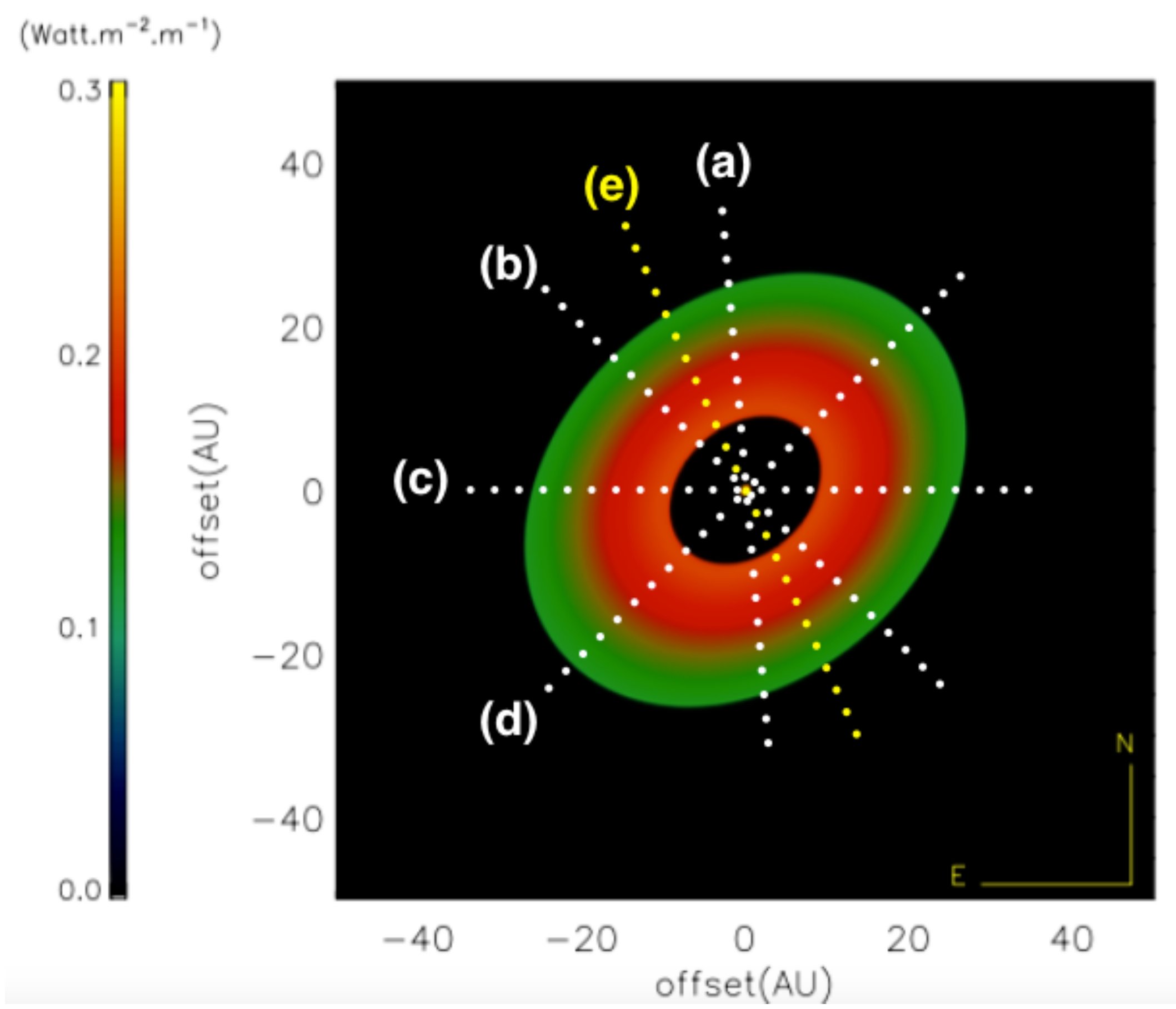} 
\end{tabular}
\caption{The bottom-right plot is the synthetic image of HD\,100546 at $\lambda$\,=\,10\,$\mu$m, using our best model. With the projected baselines indicated we derive the correlated fluxes shown in panels (a) to (e). Overplotted on the measured correlated fluxes (black) is the slope of the correlated fluxes (solid magenta) and the best model of the slope of the correlated fluxes (dashed magenta), including the inner disk and its inner rim. The blue lines represent the best model including the inner and the outer disk. The green curve in panel\,(a) is the spline function fitted to this measurement. We list the separations derived from spline fitting ($S_{o}$), full binary model ($S_{b}$), and from our two-component disk model ($S_{m} \equiv r_{in,od}$), respectively. Panel (e) corresponds to the observation in 2005 \citep{2014A&A...562A.101P}. \label{2}}
\end{figure*}

 An example of a spline fit on a correlated flux is shown as the green curve in panel\,(a) of Fig.\,\ref{2}. 

\begin{figure}
\centering
\begin{tabular}{c}
\includegraphics[clip=,width=0.99\hsize]{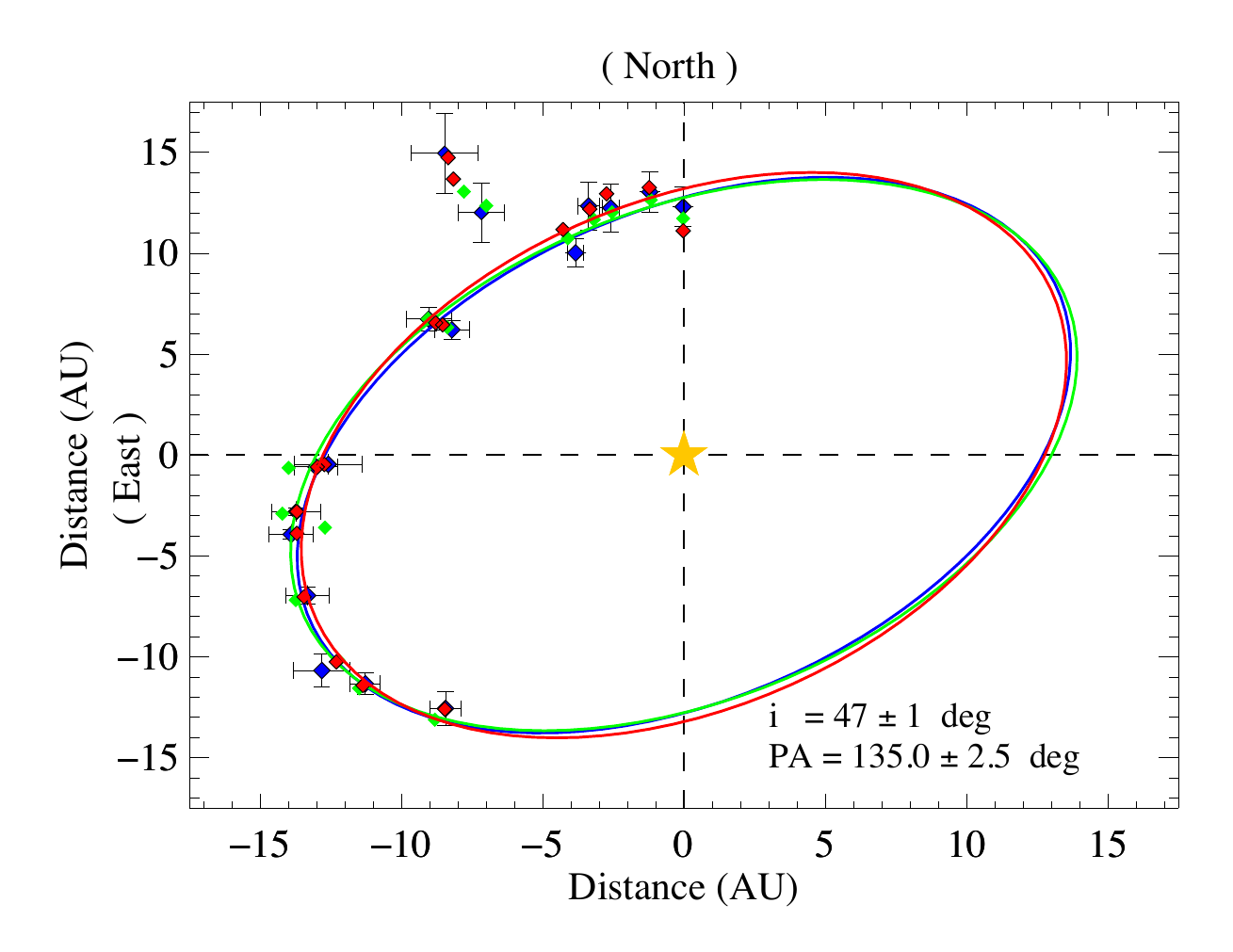} 
\end{tabular}
\caption{The separations derived from spline fitting (blue), full binary model (red), and our two-component disk model (green), respectively. The corresponding ellipse fits on these data points are also overplotted with the same color codes. The result quoted is that of the spline fitting. The two measurements that deviate from the other points have been obtained in 2004/2005 \citep{2014A&A...562A.101P}. We do not include them in our ellipse fitting.\label{ellipse}}
\end{figure}

We present in Fig.\,\ref{ellipse} the separations and their corresponding error bars as blue points. The ellipse fit to the data points gives i\,=\,47\,$\pm$\,1$^{\circ}$ and PA\,=\,135\,$\pm$\,2.5$^{\circ}$. Note that the uncertainties are only the formal ones in the framework of the specific model. We derive a radius of the ellipse of 15.7\,$\pm$\,0.8\,au. These results are roughly consistent with the values found in previous studies listed in Table\,1. Please note that the PA and the cavity size we derive is not consistent with the value derived in the MIR study by \citet{2014A&A...562A.101P}. The actual difference of the cavity size along PA of \,30\,$^{\circ}$ is 1.5\,au, i.e., the separations found for the old data are not consistent with the new data (Fig.\,\ref{ellipse}). The reason for this may be the variability of the source in the 7 year period (Sect.\,\ref{subsubsec:autonumber}).

The determination of the minima locations with the spline method, however, does not take into account another restriction of the binary model. The periodicity of the binary signal has to be consistent with a maximum at spatial frequency 0. To justify and confirm the minima determination of the correlated fluxes, we use a simplistic binary model of two point sources with equal brightness. In this model the correlated flux only depends on the angular separation along the baseline direction. The data points and the ellipse fitted to the derived separations are shown in red in Fig.\,\ref{ellipse}. We find i\,=\,46\,$\pm$\,0.5$^{\circ}$ and PA\,=\,134\,$\pm$\,1$^{\circ}$ as well as a semi-major axis of 15.3\,$\pm$\,0.7\,au. This confirms that the previous results are robust and shows that the two approaches are consistent with each other. Please note that the two older measurements (2004, 2005) represented by the red points in Fig.\,\ref{1} are not included in our geometrical model described above.

\subsection{Two-component disk model} \label{subsubsec:autonumber}

 For a more detailed interpretation of our data, we use a semi-analytical model \citep{2015Ap&SS.355..105J,2018MNRAS.473.3147J}. This model of protoplantary disks around Young Stellar Objects (YSOs) \citep{2015A&A...579A..81J} takes the temperature and surface density power law into account. Pre-transitional disks can be modeled with two components separated by a gap.

 In a first step, we start, however, with constructing an one-component disk model, which only comprises an inner rim and an inner disk at less than 1\,au both with a temperature and optical depth power law (Table\,2). The outer disk is added later. The inclination $i$ and the position angle PA of the disk are fixed to the values constrained by our ellipse fitting (on data points fitted by spline function; see Sect.\,3.1). The temperature at the inner radius of the rim ($T_{in,rim}$) in the rim temperature power law, is a free parameter, ranging from 600\,K \citep{2014A&A...562A.101P} to 1750 K \citep{2011A&A...531A...1T}. 
The exponent $q_{rim}$ varies between 0.1 to 1.5.


The temperature of the inner radius of the inner disk ($T_{in,id}$) is equal to the temperature of the inner rim at inner radius of the inner disk ($r_{in,id}$).
The exponent $q_{id}$ varies again between 0.1 to 1.5.

\begin{table}
\caption{The best parameters obtained for an inner disk and outer disk separated by a gap. The parameters marked with $a$, $b$ and $c$ are the ones constrained by the geometrical model, the one-component disk model and the two-component disk model respectively. The parameter marked with $c\,\star$ is constrained for each individual observation by our two-component-disk model with all other parameters fixed.\label{table:2}} 
\setlength{\tabcolsep}{5pt} 
\singlespace
\centering
\small

\renewcommand{\arraystretch}{0.8}
\begin{tabular}{l*{6}{c}r}
\hline
\hline
Geometry &    & \\ 
\hline
Inclination &  47\,$\pm$\,1$^{\circ }$ & $ a $\\ 
PA            &  135.0\,$\pm$\,2.5$^{\circ }$ &  $ a $\\  
 \hline
Inner rim   &   &\\ 
\hline
$T_{\rm r,rim}\,=\,T_{\rm in,rim}({r}/{r_{\rm in,rim}})^{-q_{rim}}$ &   & \\ 
$T_{\rm in,rim}$ & 1200 K&  $ b$  \\ 
$r_{in,rim}$ &  0.24\,au & \\ 
$r_{out,rim}$ &  0.34\,au & \\ 
$q_{rim}$ &  0.4 & $ b$ \\ 
$\tau_{\rm r,rim}\,=\,\tau_{\rm in,rim}({r}/{r_{\rm in,rim}})^{-p_{rim}}$ & & \\
$\tau_{\rm in,rim}$ & 0.15 &  $b $\\
$p_{rim}$ &  1.5 & $b $\\
\hline
Inner disk   &   &\\ 
\hline
$T_{\rm r,id}\,=\,T_{\rm in,id}({r}/{r_{\rm in,id}})^{-q_{id}}$ &   & \\ 
$T_{\rm in,id}\,=\,T_{\rm in,rim}({r_{\rm in,id}}/{r_{\rm in,rim}})^{-q_{rim}}$&   & \\ 
$r_{in,id}$ &  0.34 au & \\ 
$r_{out,id}$ &  0.7 au & \\ 
$q_{id}$ &  0.2 & $b $\\ 
$\tau_{\rm r,id}\,=\,\tau_{\rm in,id}({r}/{r_{\rm in,id}})^{-p_{id}}$ & & \\
$\tau_{\rm in,id}\,=\,\tau_{\rm in,rim}({r_{\rm in,id}}/{r_{\rm in,rim}})^{-p_{rim}}$ & & \\
$p_{id}$ &  1.4 & $b $\\
\hline
Outer disk   &   &\\ 
\hline
$T_{\rm r,od}\,=\,T_{\rm in,od}({r}/{r_{\rm in,od}})^{-q_{od}}$ &   & \\ 
$T_{\rm in,od}$ & 300\,K  & $c $\\ 
$r_{in,od}$ &  (10--16.5)\,au & $c\, \star$\\ 
$r_{out,od}$ &  35\,au & \\ 
$q_{od}$ &  1.5 & $c $ \\ 
$\tau_{\rm r,od}\,=\, \sin(\pi/2 \frac{(r-r_{\rm in,od})}{(r_{\rm out,od}-r_{\rm in,od})})$ & &\\ \\ 
\hline
\end{tabular} 
\end{table}       

As shown in Table\,2, we fix the rim location to the value of $r_{in,rim}$\,=\, 0.24 constrained by \citet{2010A&A...511A..75B} and \citet{2011A&A...531A...1T}. We also set the inner radius of the inner disk, $r_{in,id}$, which is equal to the outer radius of the inner rim as well as the outer radius of the inner disk to the values measured by \citet{2014A&A...562A.101P}, i.e., 0.34\,au, and 0.7\,au, respectively.  

In our disk model, the observer receives the following flux for each disk elementary surface area A, i.e., the area defined by our pixel size in the simulated brightness maps of our model :  

\begin{equation}
       dF_{\lambda}(i)\,=\,B_{\lambda} \left[  T_{\rm r}\right] \left[1-exp\left(-\frac{\tau_{\rm r}}{cos(i)}\right)\right]  \left(\frac{A}{D^{2}}\right),
\end{equation} 
  
\noindent where, $B_{\lambda}$  is the Planck function, which depends on the temperature of the disk at each radius, the quantity $A/{D^{2}}$ represents the solid angle of each elementary surface area as seen at the distance $D$, $i$ is the disk inclination and $\tau_{\rm r}$ is the optical depth as a function of radius.

We first used an optically thick rim and an inner disk where $\tau_{rim}$\,=\, $\tau_{id}$\,=\,1. Since using this definition we could not reproduce the slope of our correlated fluxes, we then assumed a power law for the optical depth, $\tau_{\rm r}\,=\,\tau_{\rm in}({r}/{r_{\rm in}})^{-p}$ for which the value decreases with increasing distance from the central star. For the inner rim, $\tau_{\rm in}$ is assumed to be $\tau_{in,rim}$ at $r_{in,rim}$. For the inner disk, $\tau_{\rm in}$ is calculated from the optical depth of the inner rim at $r_{in,id}$. The slope of the correlated fluxes are very sensitive to the value of $\tau_{\rm in}$. We vary this parameter in a range between 0 to 1 to obtain the best slope. The exponent of p in the optical depth power-law is defined as $p_{\rm rim}$ for the inner rim and $p_{\rm id}$ for the inner disk. We vary these parameters between 0.1 to 1.5.

After creating the image of the disk (including the star) for each wavelength, we calculate the correlated flux. The correlated flux is the modulus of the Fourier transform for the observed spatial frequency. The inner disk in our model only creates a slope with which we can reproduce the slope of our data. To do so, we first define the slope of each correlated flux using the lower and the upper wavelengths and their corresponding correlated flux value as shown in Fig.\,\ref{2} and Fig.\,4 by the solid magenta lines. Exploring the range of free parameters in our model, we could reproduce the slope of the correlated fluxes assuming an inner disk smaller than 1\,au. The reduced $\chi^{2}$ of our best model, which is the difference between the model and the data weighted by the number of elements of the data is 1.4. We summarize the parameters constrained in our best model marked with $b$ in Table\,2. We present the best model by dashed magenta lines in Fig.\,\ref{2} and Fig.\,4.

In a next step, we use a two-component disk model adding an outer disk to the inner one, separated by a gap. The temperature of the inner radius of the outer disk ($T_{in,od}$) in the outer disk temperature power law is a free parameter and varies in a physical consistent temperature range 200--300\,K as used for an illuminated wall by \citet{2011A&A...531A..93M}. The exponent $q_{od}$ in the power law varies between 0.1 to 1.5. We set the outer radius of the outer disk {\bf ($r_{out,od}$)} to a high enough value of 35\,au to provide a gradual decrease in the radial brightness, as doing otherwise may introduce spurious interferometric signal. For the optical depth we use the expression introduced by \citet{2014A&A...562A.101P}. Since the contrast between the wall of the outer disk and the gap causes a high contrast in the flux, they introduced a sine function (see Table\,2) to gradually smooth the peak of the flux behind $r_{in,od}$. With respect to the inner disk we kept the best parameters obtained for the one-component disk model reproducing the slope of our correlated fluxes (see Table\,2). 

The temperature of the outer disk and the exponent $q$ affect the level of the correlated fluxes and the inner radius of the outer disk $r_{in,od}$ affects the location of their minima. Therefore, we first adjust the level of the correlated fluxes for all observations exploring the range of the temperature and $q$ for a given $r_{in,od}$. Then, we vary $r_{in,od}$ for each observation between 7 to 17\,au assuming that the other parameters presented in Table\,\ref{table:2} are fixed. The results vary between 10 and 16.5\,au. This step is done manually. The correlated fluxes are presented in Fig.\,\ref{2} for five different baseline configurations. The remaining plots are presented in Fig.\,\ref{alldata}. Adding the outer disk to the one-component disk model causes the wavy pattern of our correlated fluxes. Using our best model, we successfully reproduce the minima separations and roughly the level of each correlated flux for the specific value of $r_{in,od}$ for the given distance. The data points and the corresponding fitted ellipse are represented in Fig.\,\ref{ellipse}. The under predicted flux in our model compared to some of the observed data in Fig.\,\ref{2} and Fig.\,\ref{alldata} could be due to an asymmetry at the corresponding orientations, which has not been included in our symmetrical model. 

We see also a pronounced signal in the differential phases, which is produced by an asymmetry in the outer edge of the gap as detected by \citet{2014A&A...562A.101P}. This asymmetry may be caused by a bright wall of the outer disk, which is tilted away from the observer. We find a flux ratio of less than 0.25 in the observed differential phases. We will discuss this issue in a forthcoming paper in detail.

Indeed, the separations as derived from our data are representative of the geometrical structure of the disk and allow a unique estimation of the gap size. As demonstrated in Figs.\,\ref{2} \& \ref{alldata} we can reproduce the minima locations of the correlated fluxes for the older measurements. However, the corresponding separations deviate from the ones found from our new data. This implies that the deviation of the separation measurements shown in Fig.\,2 between the old data and new data could be due to the difference in the geometry of the disk, i.e., the projected gap size in that direction. We note that \citet{2014A&A...562A.101P} already witnessed variability even between the two older data points, which had roughly the same baseline length and orientation and were taken with only about one year time difference. They suggested that this time variability is because of the different spatial distribution of the brightness on AU-scales. We expected to observe the variability due to the 7\,years time difference between the first two epochs and our new data, which was confirmed by our new cavity size. The exact reason for this variability, however, requires further investigations and might be a combination of temperature and strcutural effects.

\section{Summary and conclusion}

We have presented multi-epoch interferometric N-band data with a good uv-coverage obtained with VLTI/MIDI. The measured correlated fluxes are characterized by a wavy pattern, which varies with baseline length and orientation. However, some minima are less significant than others, which may or may not be related to the silicate feature that is present in the total N band fluxes \citep{2005A&A...437..189V}. Our multi-epoch data allows us to estimate the gap size of the disk of HD\,100546 in the N band. We show that the separation obtained by direct measurement of the separation of the minima of our new data in 2012 and 2013 are satisfactorily fitted by an ellipse. We derive an inclination of 47\,$\pm$\,1$^{\circ }$ and a PA of 135\,$\pm$\,2.54$^{\circ}$, which are consistent with the values estimated by previous studies. We also constrain gap size to 15.7\,$\pm$\,0.8\,au. We show by our model that the inner radius of the outer disk estimated for each observation is consistent with the separation obtained from the geometrical model. Please note that our projected gap size estimation takes into account the updated distance of HD\,100546 measured by GAIA, $\sim$\,13$\%$ larger than the distance of 97\,pc commonly used in literature. Taking into account this fact, our projected gap size estimation bears striking similarity to the multi-wavelengths values found by previous studies listed in Table\,1. However, one has to be careful in comparing the sizes, since the different wavelength regimes and observing methods trace potential different structures.

Reconstructing images of the disk of HD\,100546 at L and N band with the upcoming second-generation VLTI instrument MATISSE \citep[the Multi AperTure mid-Infrared SpectroScopic Experiment,][]{2006SPIE.6268E..0ZL} will constitute a unique perspective to further assess the nature of the geometrical structure of HD\,100546 and will be a natural continuation for our study. MATISSE will recombine up to four telescopes. More detailed models including also our data obtained with the Auxiliary Telescopes (ATs) will be presented in a forthcoming paper.

\acknowledgments
The research leading to these results has received funding from the European Union's Horizon 2020 research and innovation programme under Grant Agreement 730890 (OPTICON). O. Pani\'c is funded by the Royal Society Dorothy Hodgkin Fellowship. Based on observations collected at the European Organization for Astronomical Research in the Southern Hemisphere.

\appendix
\section{Additional online material}

\clearpage

\begin{table*}
\centering
\label{table:3} 
\caption{Log of observations. The excluded calibrator stars are indicated with an asterisk.}
\begin{tabular}{c c c c c c c} 
\hline\hline
 \textbf{Program IDs} &\textbf{Tel.}& \textbf{Date} & \textbf{UT start} & \textbf{Target} & \textbf{$B_{p}$[m]}& \textbf{PA$^{\circ}$}  \\ [0.5ex] 
\hline
 090.C-0931(A) &U1-U3&2012--12--31&05:29:10 & HD\,94452& 94.8 & 3.7 \\ 
 090.C-0931(A) &U1-U3&2012--12--31&05:48:52 & HD\,100546& 88.4 & 0.1 \\ 
 090.C-0931(A) &U1-U3&2012--12--31&06:05:45 & HD\,102461& 97.5 & 0.6 \\ 
 \hline
090.C-0931(B) &U1-U3&2012--12--31&06:26:43 &HD\,98292 & 90.1 & 9.9  \\ 
 090.C-0931(B) &U1-U3&2012--12--31&06:52:18 & HD\,100546 & 87.9 & 11.9 \\ 
090.C-0931(B) &U1-U3&2012--12--31&07:07:26 & HD\,102839  $\star$ & 87.9 & 11.7 \\  
 \hline
 090.C-0931(B) &U1-U3&2013--03--01&01:37:42 & HD\,94452& 94.8 & 4.5 \\
 090.C-0931(B) &U1-U3&2013--03--01&02:20:39 &HD\,100546 & 88.3 & 5.3 \\ 
 090.C-0931(B) &U1-U3&2013--03--01&02:37:08 &HD\,102461 & 97.4 & 5.2 \\
 \hline
090.C-0931(B) &U1-U3&2013--03--01&02:53:17 &HD\,98292 & 89.7 & 14.0\\ 
090.C-0931(B) &U1-U3&2013--03--01&03:14:26 &HD\,100546 & 87.5 & 15.3\\ 
090.C-0931(B) &U1-U3&2013--03--01&03:29:24 &HD\,102839 & 87.5 & 15.0\\ 
 \hline 
 090.C-0931(E) &U1-U4&2013--03--01&04:12:01 &HD\,98292 & 126.3 & 52.2 \\
 090.C-0931(E) &U1-U4&2013--03--01&04:28:49 & HD\,100546& 125.5 & 52.8 \\ 
 090.C-0931(E) &U1-U4&2013--03--01&04:48:41 & HD\,102839& 125.3 & 53.6 \\ 
 \hline
 090.C-0931(C) &U3-U4&2013--03--01&05:25:12 & HD\,98292& 60.7 & 106.1\\
090.C-0931(C) &U3-U4&2013--03--01&05:40:55 & HD\,100546& 60.5 & 105.8\\ 
090.C-0931(C) &U3-U4&2013--03--01&05:58:27 & HD\,102491& 61.4 & 108.9\\ 
\hline 
 090.C-0931(C) &U3-U4&2013--03--01&07:05:13 & HD\,98292& 62.2 & 128.7 \\
 090.C-0931(C) &U3-U4&2013--03--01&07:27:08 &HD\,100546 & 62.1 & 129.8 \\ 
 \hline
090.C-0931(C) &U3-U4&2013--03--01&07:37:21 & HD\,100546& 62.1 & 132.1\\
090.C-0931(C) &U3-U4&2013--03--01&07:57:18 & HD\,102839& 62.1 & 132.9\\ 
\hline 
 090.C-0931(C) &U3-U4&2013--03--01&06:14:59 & HD\,98292& 61.6 & 117.3 \\ 
 090.C-0931(C) &U3-U4&2013--03--01&06:33:50 &HD\,100546 & 61.4 & 117.6 \\
 090.C-0931(C) &U3-U4&2013--03--01&06:49:13 &HD\,102839 & 61.4 & 117.3 \\ 
 \hline
 090.C-0931(E) &U1-U4&2013--03--03&01:33:55 & HD\,94452& 130.2 & 28.2 \\ 
 090.C-0931(E) &U1-U4&2013--03--03&01:48:16 & HD\,100546& 128.8 & 20.9 \\ 
 090.C-0931(E) &U1-U4&2013--03--03&02:12:00 &HD\,102461 & 130.1 & 24.3 \\ 
 \hline
090.C-0931(D) &U2-U4&2013--03--03&02:58:09 & HD\,98292& 89.0 & 53.1\\ 
090.C-0931(D) &U2-U4&2013--03--03&03:13:30 & HD\,100546& 89.2 & 52.9\\ 
090.C-0931(D) &U2-U4&2013--03--03&03:35:27 &102461  & 87.8 & 54.8\\ 
  \hline
 090.C-0931(D) &U2-U4&2013--03--03&05:41:19 &HD\,98292 & 88.5 & 89.9 \\
 090.C-0931(D) &U2-U4&2013--03--03&06:05:19 &HD\,100546 & 88.3 & 92.1 \\
 090.C-0931(D) &U2-U4&2013--03--03&06:23:43  & HD\,102839 $\star$& 88.3 & 92.5 \\
 \hline
090.C-0931(D) &U1-U4&2013--03--03&06:56:03 &HD\,98292 & 113.8 & 91.3\\ 
090.C-0931(D) &U1-U4&2013--03--03&07:11:13 & HD\,100546& 113.4 & 92.6\\ 
090.C-0931(D) &U1-U4&2013--03--03&07:31:07 &HD\,102839 $\star$ & 113.6 & 93.5\\
  \hline
 090.C-0931(D) &U2-U4&2013--03--03&09:21:31 & HD\,102839& 82.5 & 138.5 \\
 090.C-0931(D) &U2-U4&2013--03--03&09:36:52 & HD\,100546& 81.3 & 146.1 \\ 
 090.C-0931(D) &U2-U4&2013--03--03&09:55:47 &HD\,120404 & 85.5 & 145.2 \\
  \hline
090.C-0931(E) &U1-U4&2013--03--04&07:26:35 & HD\,98292& 109.0 & 99.4\\  
090.C-0931(E) &U1-U4&2013--03--04&07:42:16 & HD\,100546& 109.7 & 101.5\\  
090.C-0931(E) &U1-U4&2013--03--04&08:08:23 &HD\,102839 & 108.5 & 103.2\\[1ex]  
     \hline
\end{tabular}
\end{table*}

\clearpage


\begin{figure*}
\centering
\begin{tabular}{c}
\includegraphics[clip=,width=0.3\hsize]{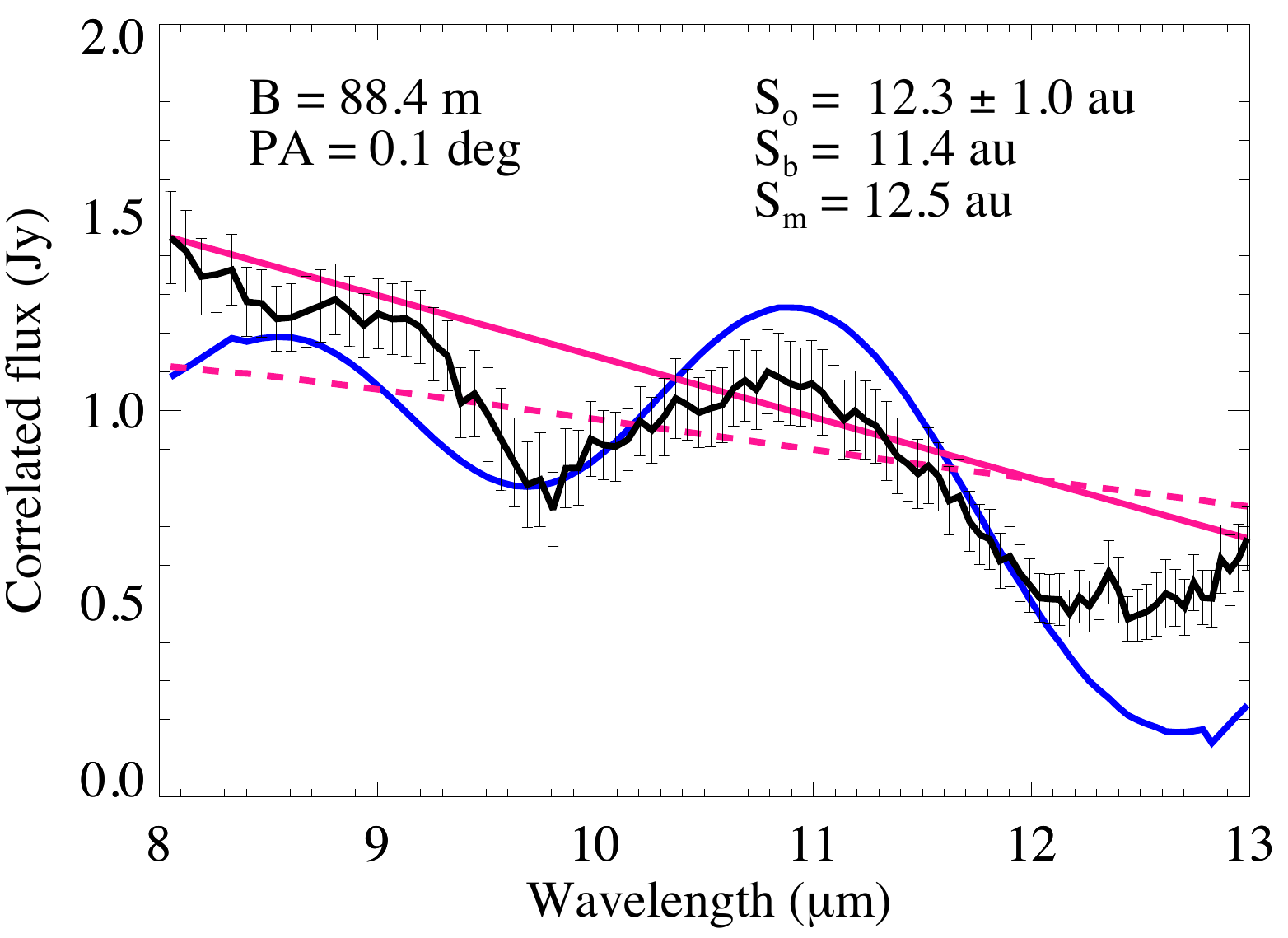} 
\includegraphics[clip=,width=0.3\hsize]{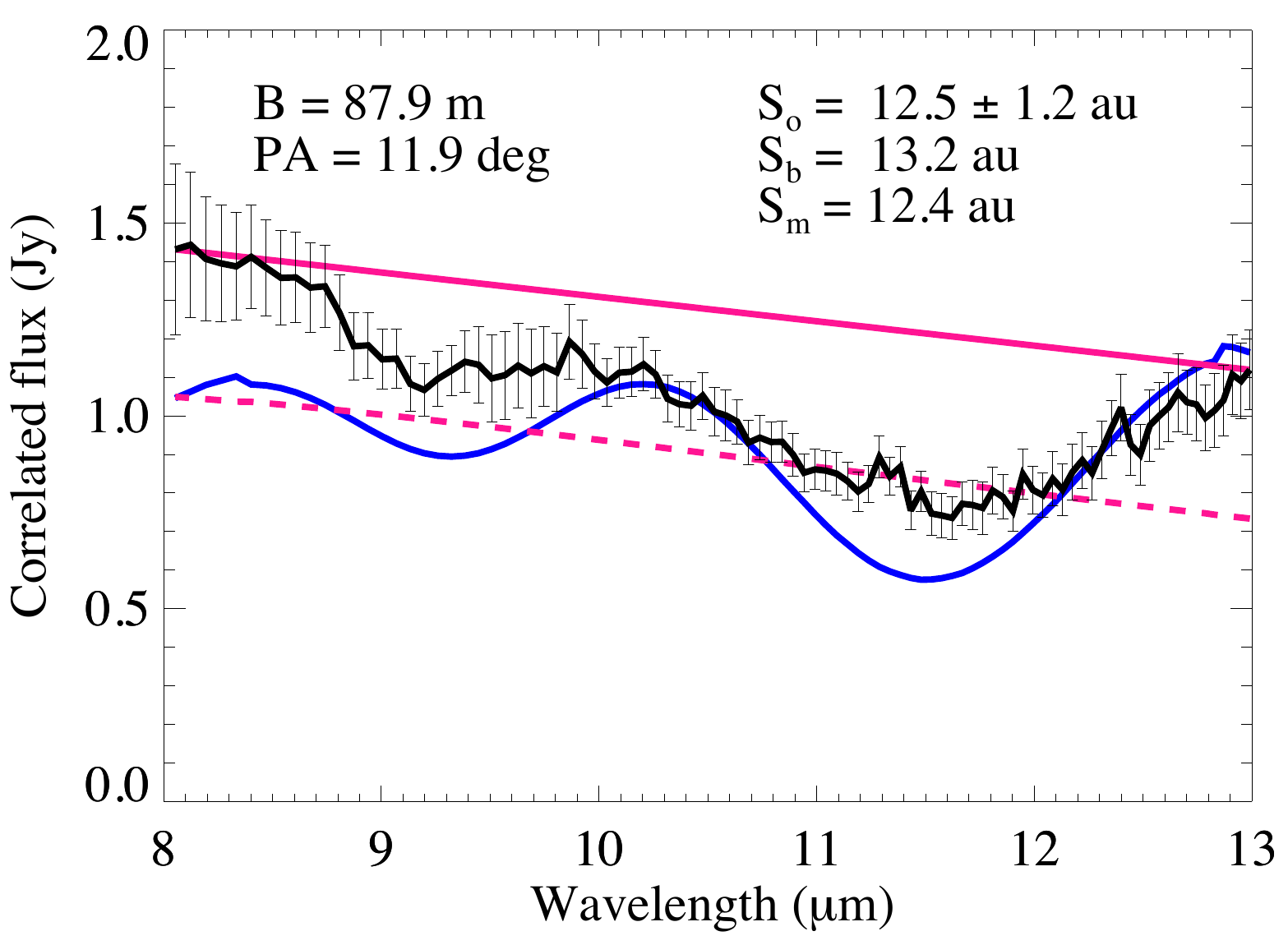}
\includegraphics[clip=,width=0.3\hsize]{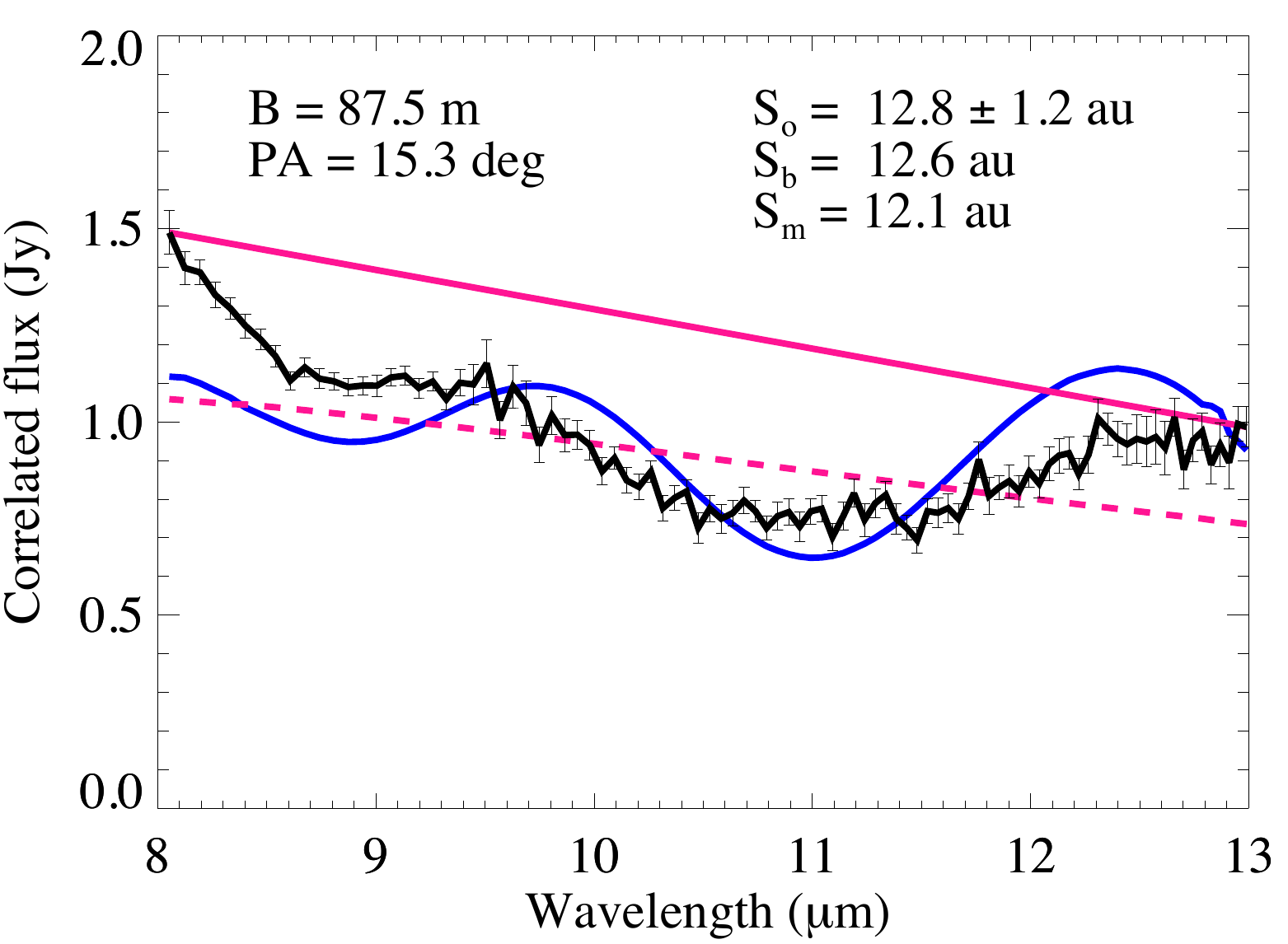}\\ 
\includegraphics[clip=,width=0.29\hsize]{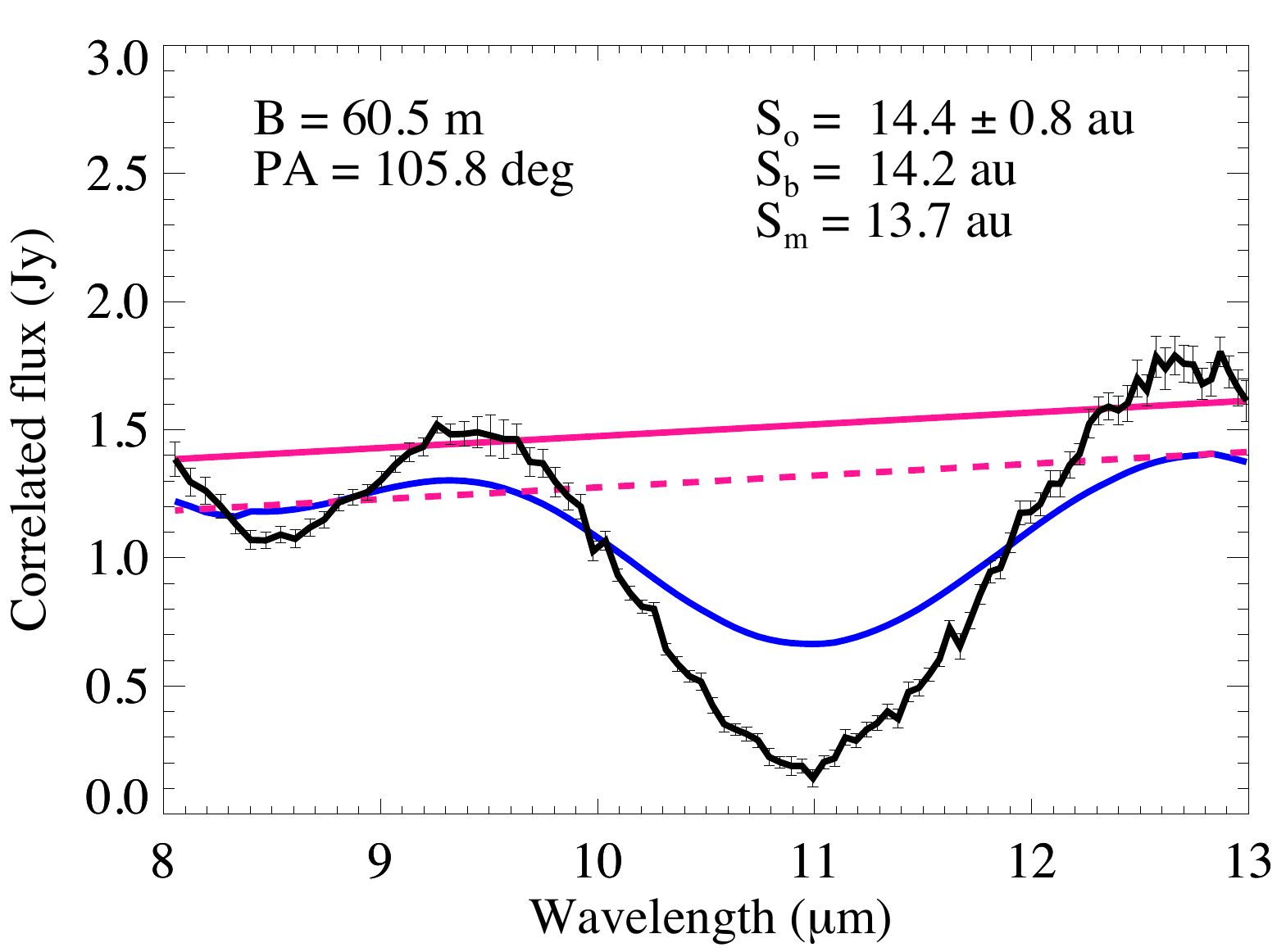}
\includegraphics[clip=,width=0.29\hsize]{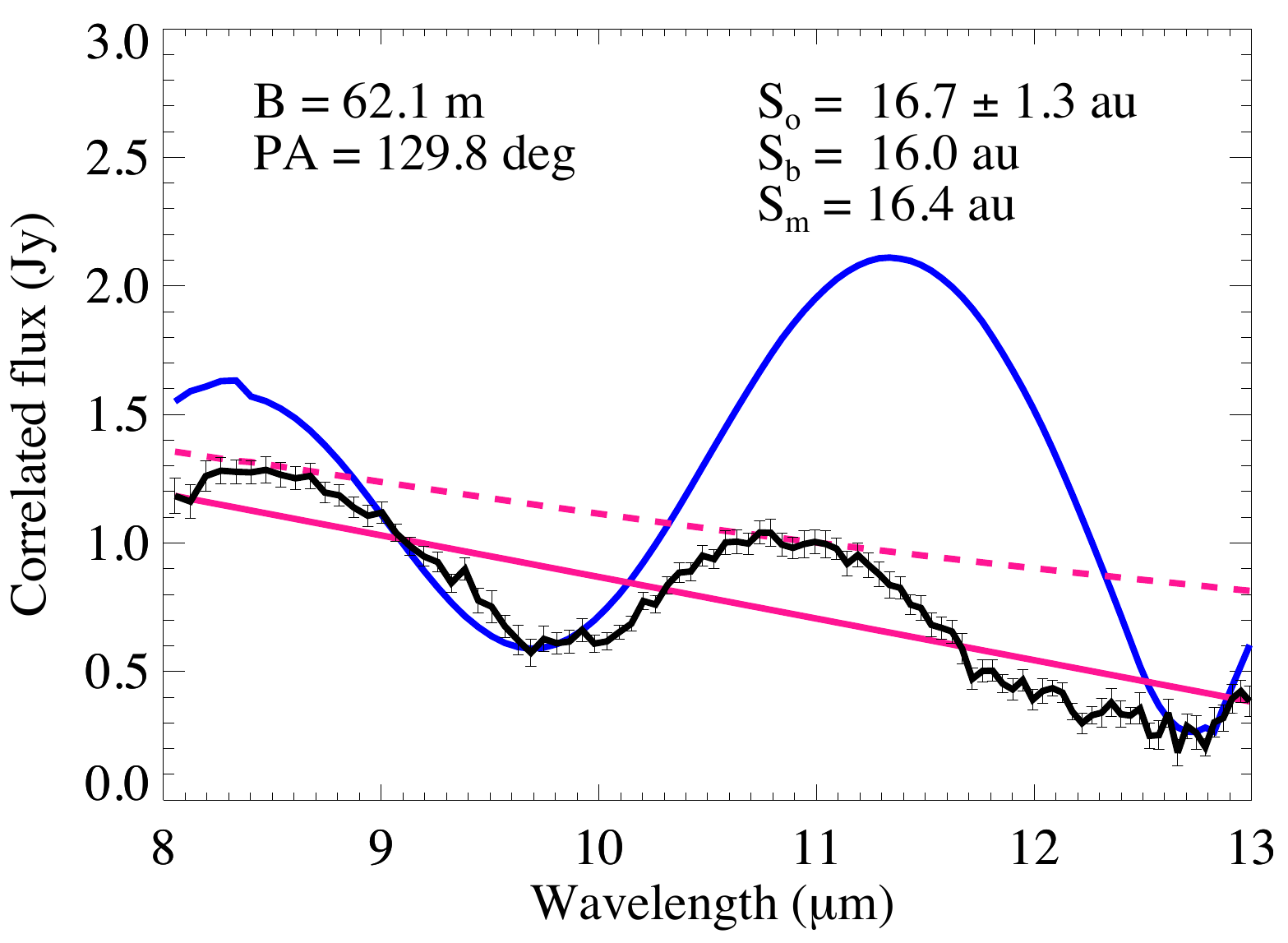}
\includegraphics[clip=,width=0.29\hsize]{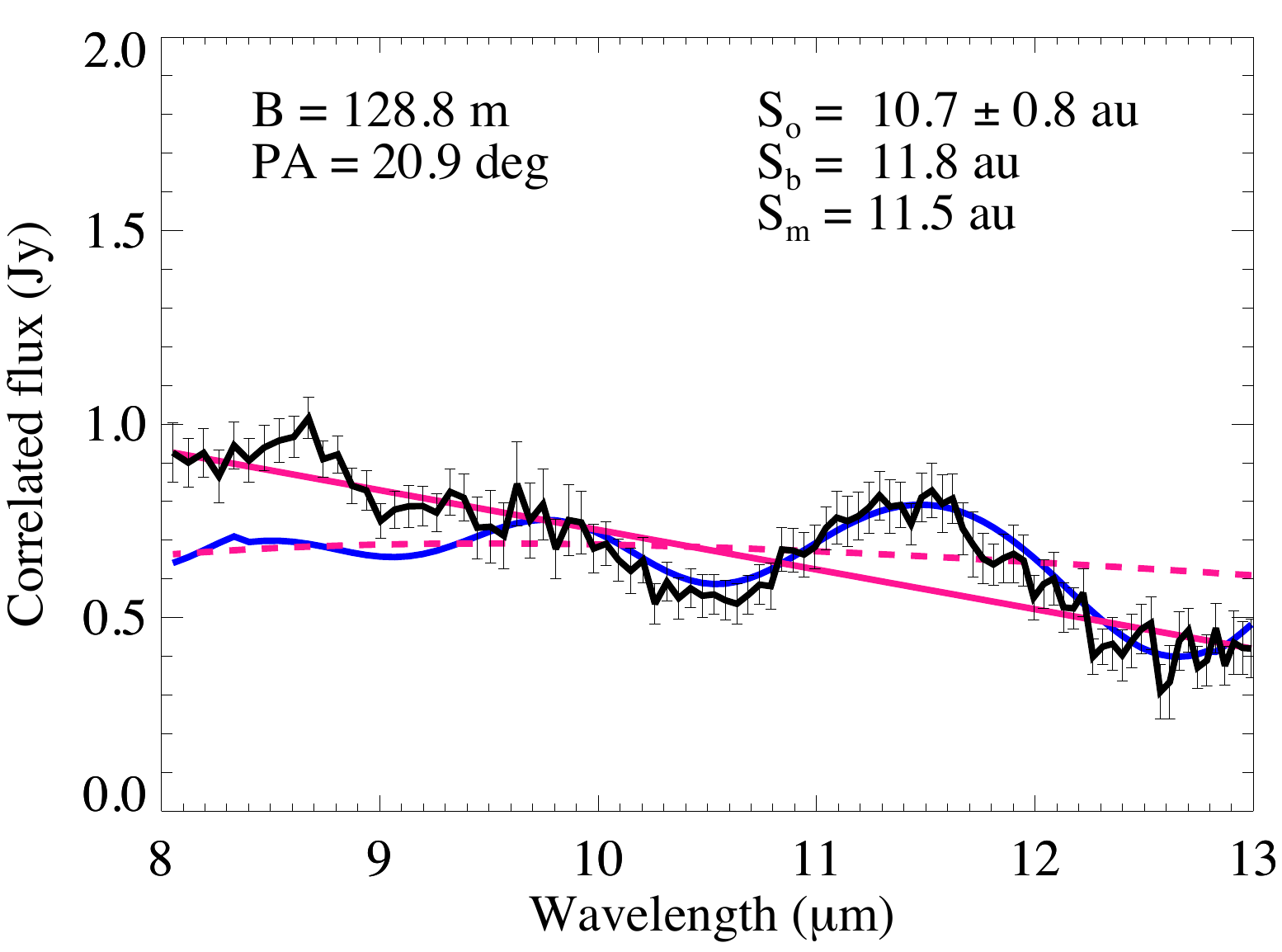}\\
\includegraphics[clip=,width=0.29\hsize]{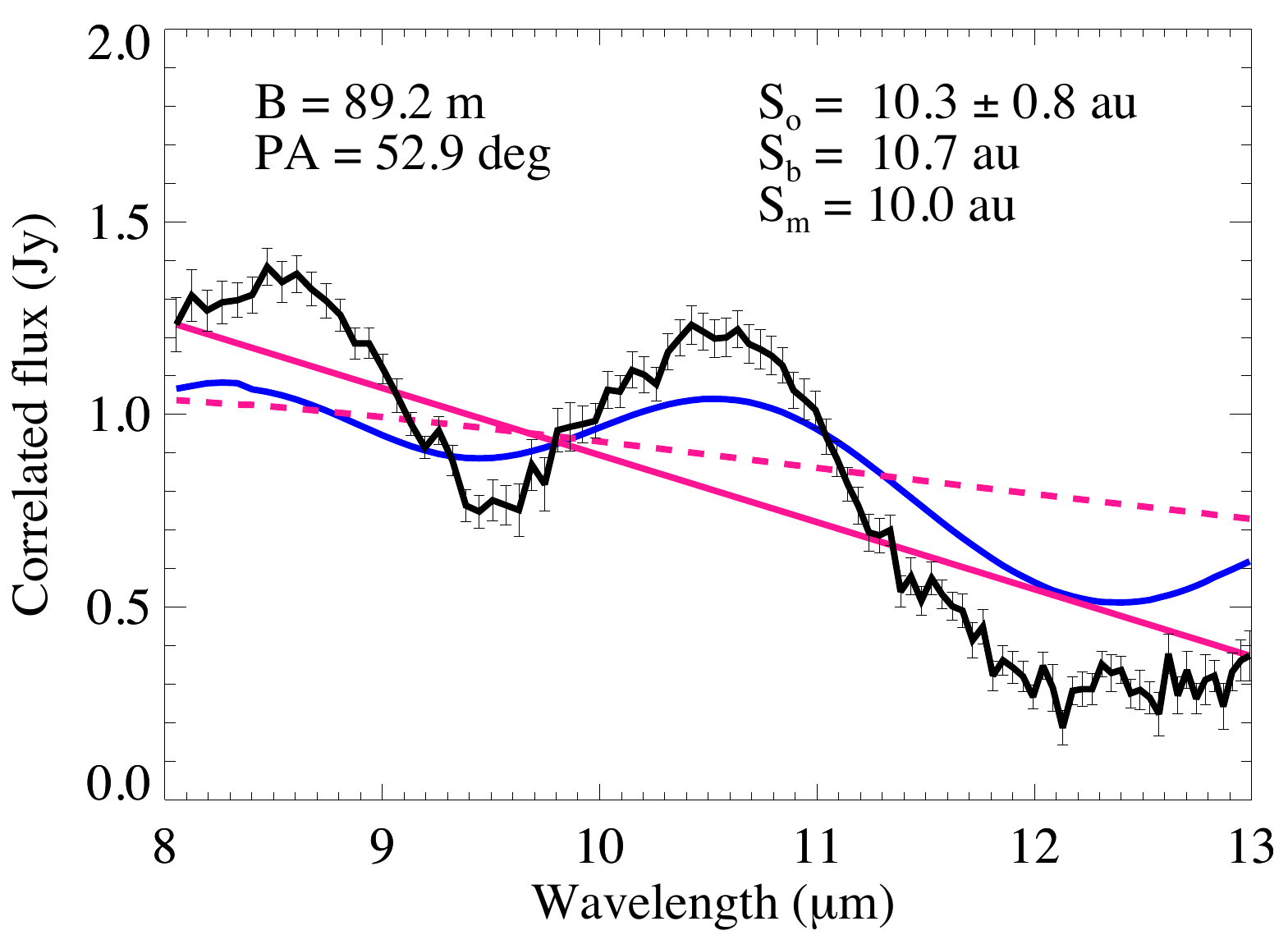}
\includegraphics[clip=,width=0.29\hsize]{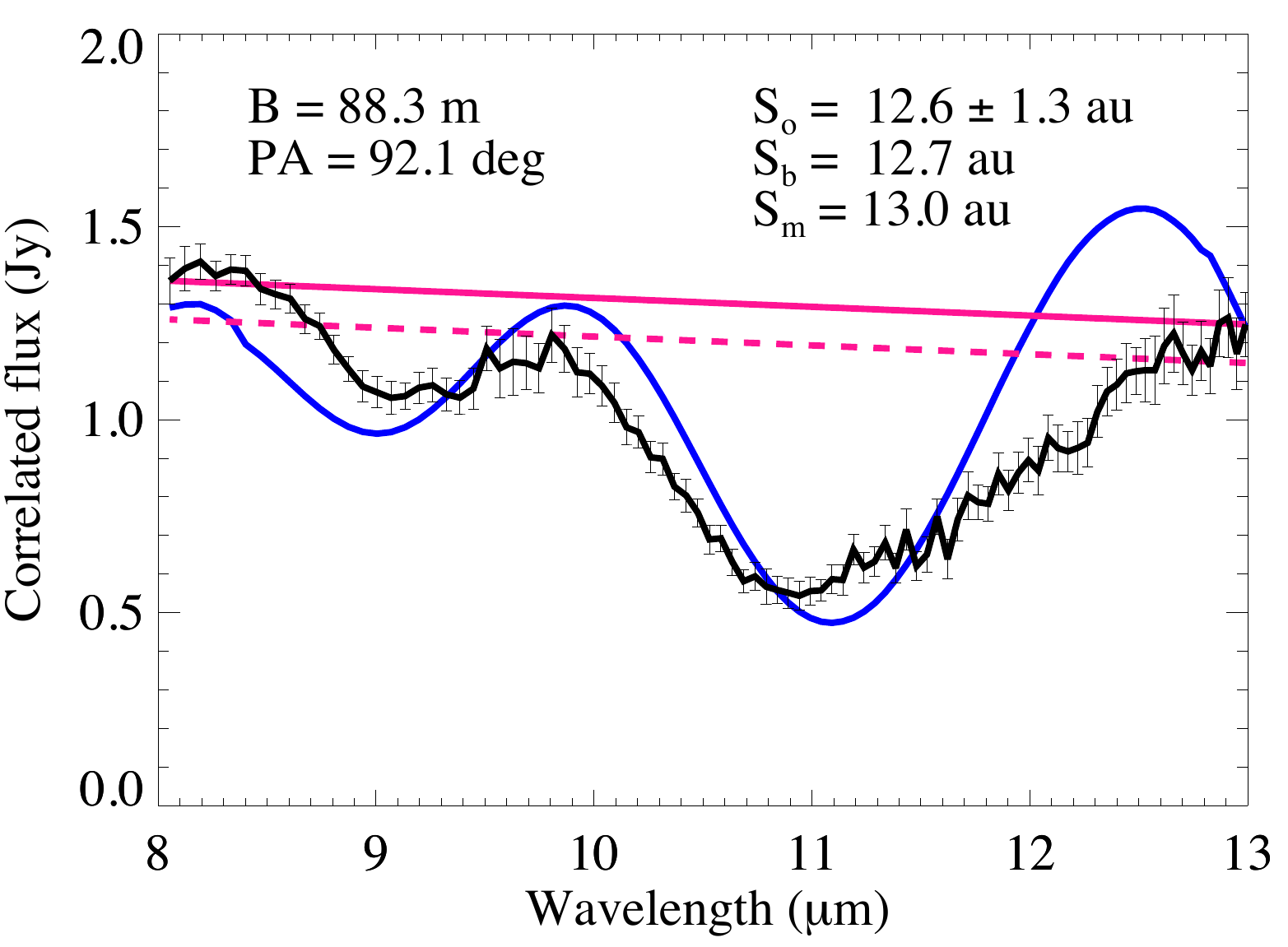}
\includegraphics[clip=,width=0.29\hsize]{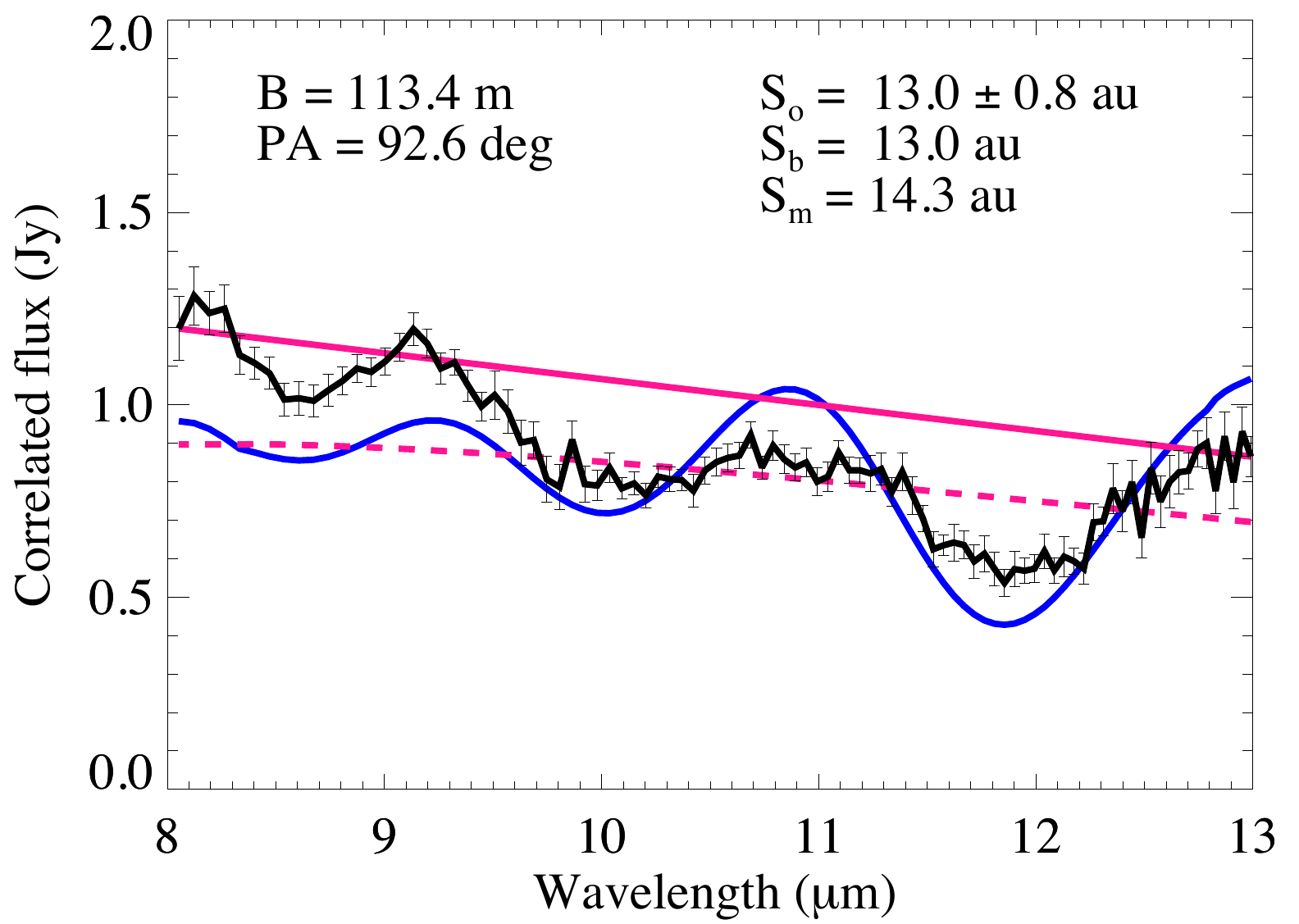}\\
\includegraphics[clip=,width=0.29\hsize]{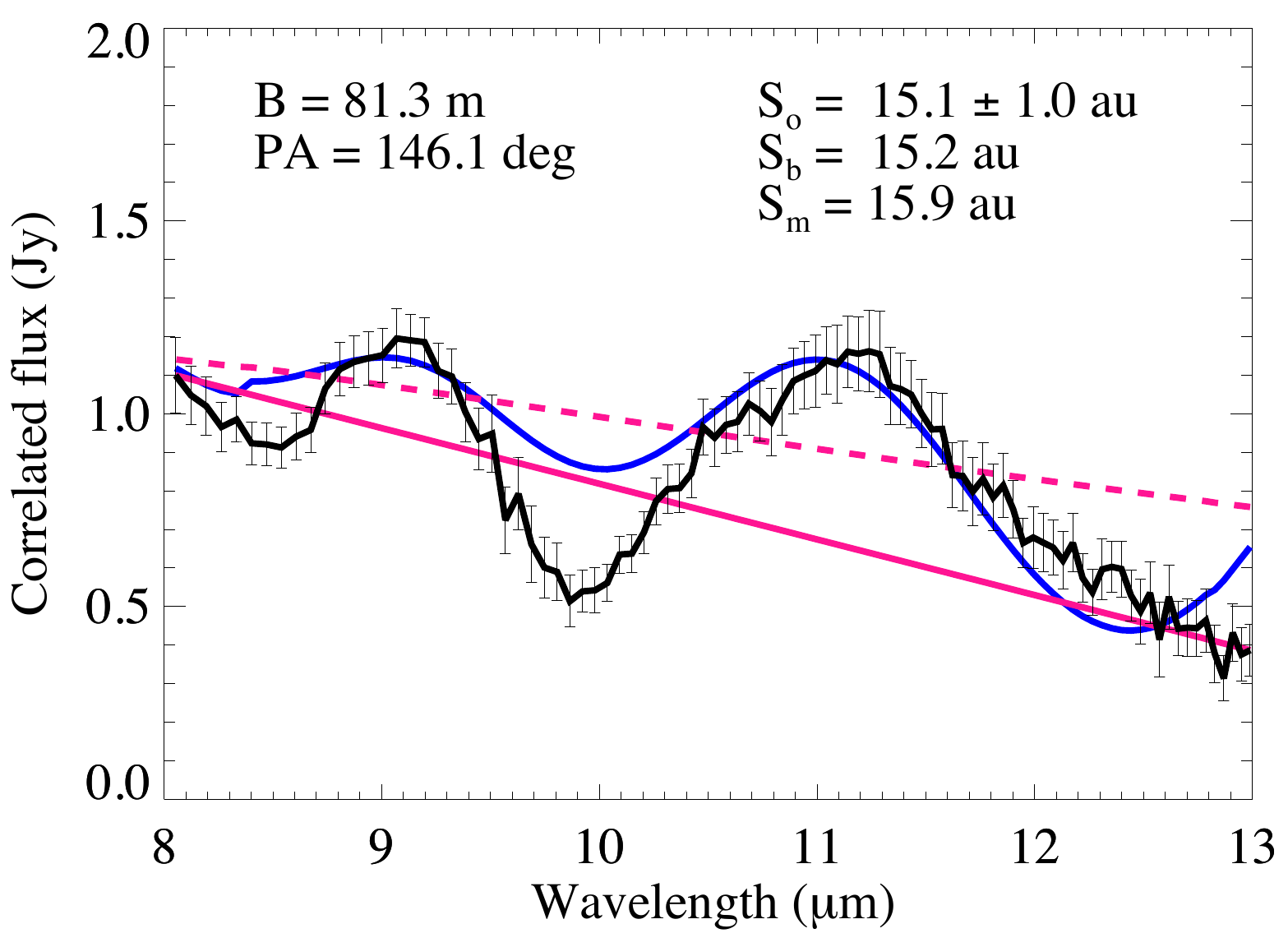}
\includegraphics[clip=,width=0.29\hsize]{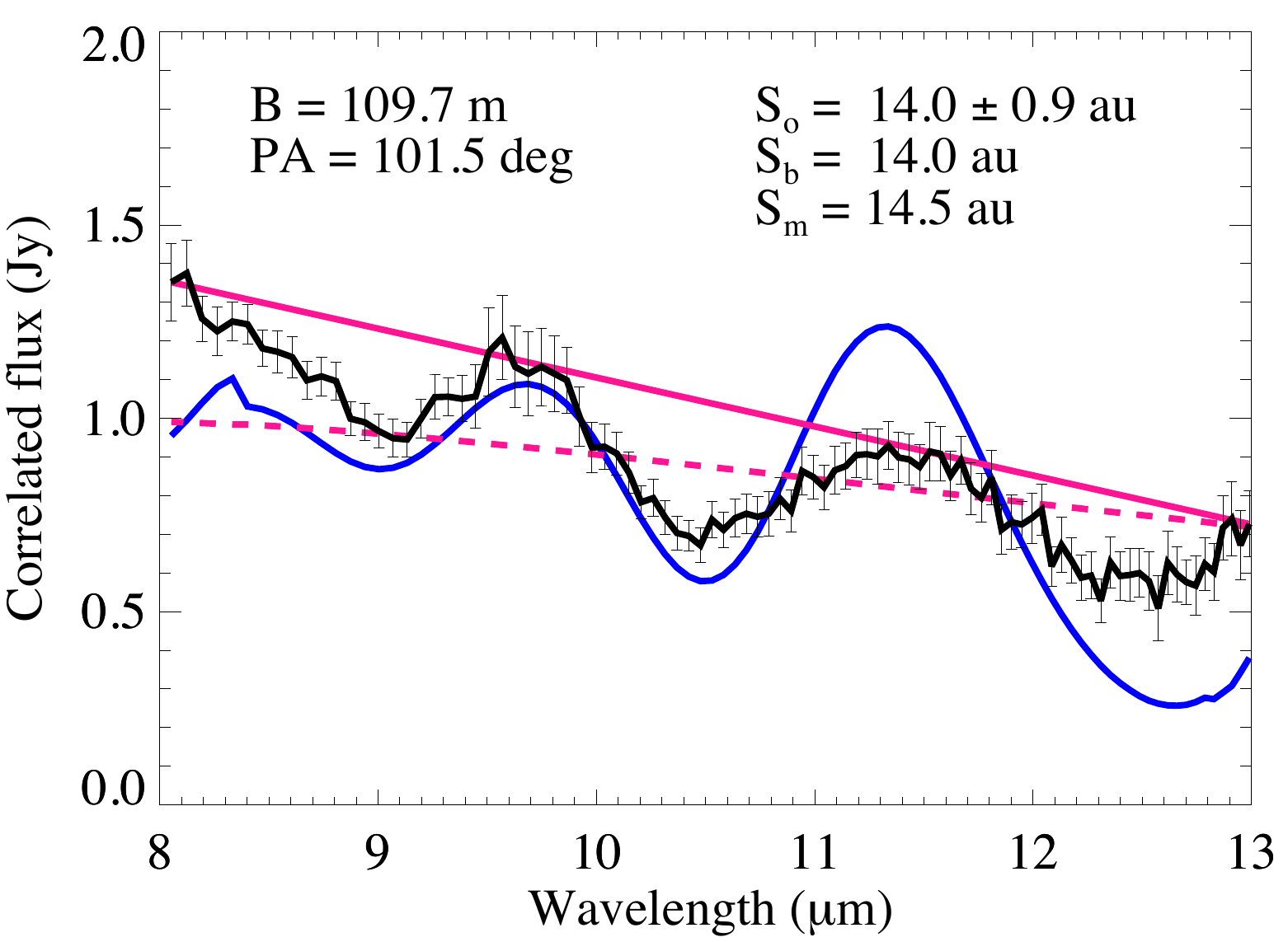}
\includegraphics[clip=,width=0.29\hsize]{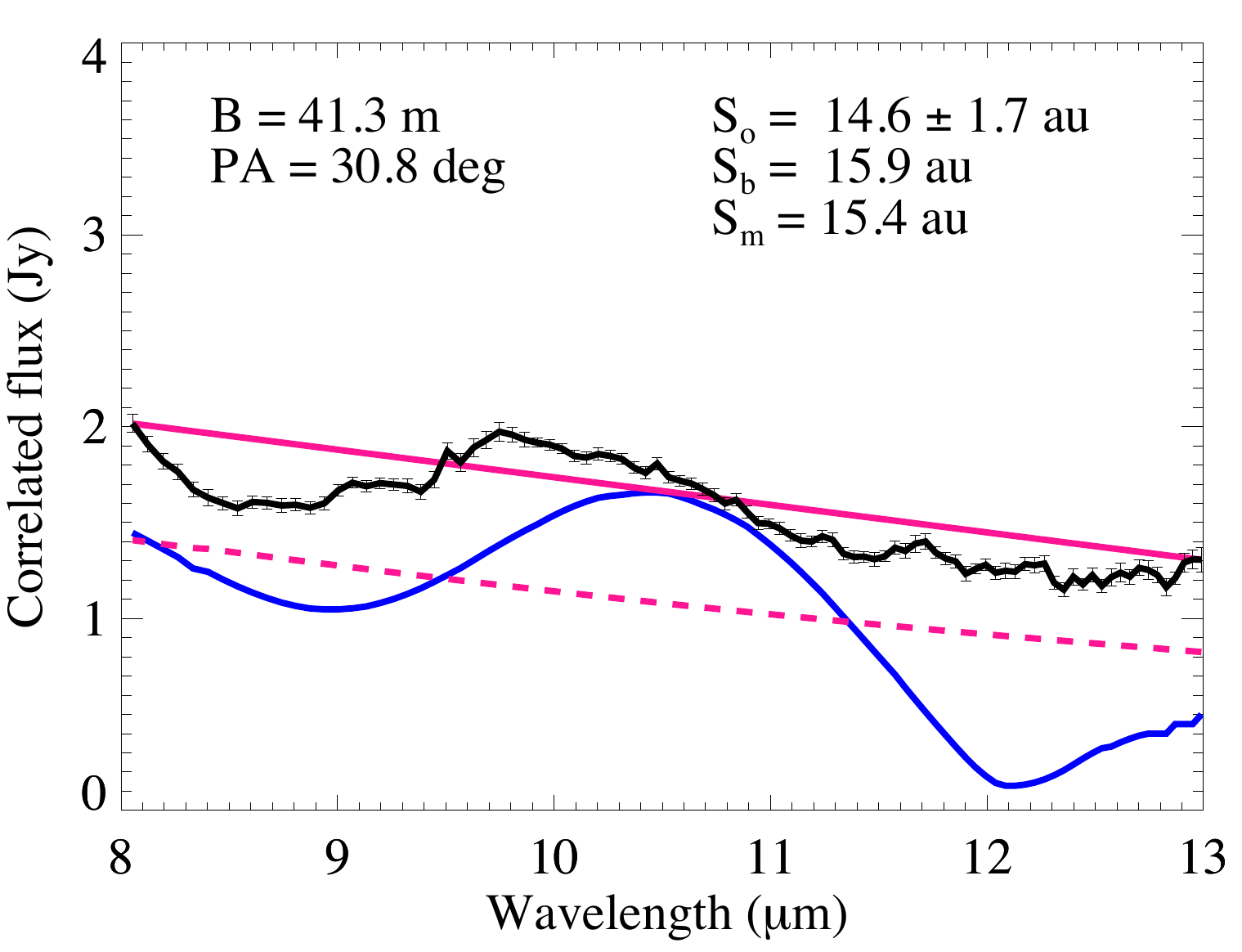}\\
\end{tabular}
\caption{Results for all measured correlated fluxes not shown in Fig.\ref{2}. In bottom-right, we use the UT data from 2004 published by \citet{2014A&A...562A.101P}. For details see Fig.\ref{2}. \label{alldata}}
\end{figure*}

\clearpage

\bibliographystyle{aasjournal}
\bibliography{Bibliography}

\begin{thebibliography}{}
\expandafter\ifx\csname natexlab\endcsname\relax\def\natexlab#1{#1}\fi
\providecommand{\url}[1]{\href{#1}{#1}}

\bibitem[{{Alexander} {et~al.}(2014){Alexander}, {Pascucci}, {Andrews},
  {Armitage}, \& {Cieza}}]{2014prpl.conf..475A}
{Alexander}, R., {Pascucci}, I., {Andrews}, S., {Armitage}, P., \& {Cieza}, L.
  2014, Protostars and Planets VI, 475

\bibitem[{{Alexander} {et~al.}(2006){Alexander}, {Clarke}, \&
  {Pringle}}]{2006MNRAS.369..229A}
{Alexander}, R.~D., {Clarke}, C.~J., \& {Pringle}, J.~E. 2006, \mnras, 369, 229

\bibitem[{{Armitage}(2011)}]{2011ARA&A..49..195A}
{Armitage}, P.~J. 2011, \araa, 49, 195

\bibitem[{{Avenhaus} {et~al.}(2014){Avenhaus}, {Quanz}, {Meyer}, {Brittain},
  {Carr}, \& {Najita}}]{2014ApJ...790...56A}
{Avenhaus}, H., {Quanz}, S.~P., {Meyer}, M.~R., {et~al.} 2014, \apj, 790, 56

\bibitem[{{Benisty} {et~al.}(2010){Benisty}, {Tatulli}, {M{\'e}nard}, \&
  {Swain}}]{2010A&A...511A..75B}
{Benisty}, M., {Tatulli}, E., {M{\'e}nard}, F., \& {Swain}, M.~R. 2010, \aap,
  511, A75

\bibitem[{{Bouwman} {et~al.}(2003){Bouwman}, {de Koter}, {Dominik}, \&
  {Waters}}]{2003A&A...401..577B}
{Bouwman}, J., {de Koter}, A., {Dominik}, C., \& {Waters}, L.~B.~F.~M. 2003,
  \aap, 401, 577

\bibitem[{{Brittain} {et~al.}(2014){Brittain}, {Carr}, {Najita}, {Quanz}, \&
  {Meyer}}]{2014ApJ...791..136B}
{Brittain}, S.~D., {Carr}, J.~S., {Najita}, J.~R., {Quanz}, S.~P., \& {Meyer},
  M.~R. 2014, \apj, 791, 136

\bibitem[{{Chesneau} {et~al.}(2006){Chesneau}, {Collioud}, {De Marco}, {Wolf},
  {Lagadec}, {Zijlstra}, {Rothkopf}, {Acker}, {Clayton}, \&
  {Lopez}}]{2006A&A...455.1009C}
{Chesneau}, O., {Collioud}, A., {De Marco}, O., {et~al.} 2006, \aap, 455, 1009

\bibitem[{{Crida} \& {Morbidelli}(2007)}]{2007MNRAS.377.1324C}
{Crida}, A., \& {Morbidelli}, A. 2007, \mnras, 377, 1324

\bibitem[{{Currie} {et~al.}(2015){Currie}, {Cloutier}, {Brittain}, {Grady},
  {Burrows}, {Muto}, {Kenyon}, \& {Kuchner}}]{2015ApJ...814L..27C}
{Currie}, T., {Cloutier}, R., {Brittain}, S., {et~al.} 2015, \apjl, 814, L27

\bibitem[{{Espaillat} {et~al.}(2007){Espaillat}, {Calvet}, {D'Alessio},
  {Hern{\'a}ndez}, {Qi}, {Hartmann}, {Furlan}, \&
  {Watson}}]{2007ApJ...670L.135E}
{Espaillat}, C., {Calvet}, N., {D'Alessio}, P., {et~al.} 2007, \apjl, 670, L135

\bibitem[{{Espaillat} {et~al.}(2014){Espaillat}, {Muzerolle}, {Najita},
  {Andrews}, {Zhu}, {Calvet}, {Kraus}, {Hashimoto}, {Kraus}, \&
  {D'Alessio}}]{2014prpl.conf..497E}
{Espaillat}, C., {Muzerolle}, J., {Najita}, J., {et~al.} 2014, Protostars and
  Planets VI, 497

\bibitem[{{Follette} {et~al.}(2017){Follette}, {Rameau}, {Dong}, {Pueyo},
  {Close}, {Duch{\^e}ne}, {Fung}, {Leonard}, {Macintosh}, {Males}, {Marois},
  {Millar-Blanchaer}, {Morzinski}, {Mullen}, {Perrin}, {Spiro}, {Wang},
  {Ammons}, {Bailey}, {Barman}, {Bulger}, {Chilcote}, {Cotten}, {De Rosa},
  {Doyon}, {Fitzgerald}, {Goodsell}, {Graham}, {Greenbaum}, {Hibon}, {Hung},
  {Ingraham}, {Kalas}, {Konopacky}, {Larkin}, {Maire}, {Marchis}, {Metchev},
  {Nielsen}, {Oppenheimer}, {Palmer}, {Patience}, {Poyneer}, {Rajan},
  {Rantakyr{\"o}}, {Savransky}, {Schneider}, {Sivaramakrishnan}, {Song},
  {Soummer}, {Thomas}, {Vega}, {Wallace}, {Ward-Duong}, {Wiktorowicz}, \&
  {Wolff}}]{2017AJ....153..264F}
{Follette}, K.~B., {Rameau}, J., {Dong}, R., {et~al.} 2017, \aj, 153, 264

\bibitem[{{Gaia Collaboration} {et~al.}(2018){Gaia Collaboration}, {Brown},
  {Vallenari}, {Prusti}, {de Bruijne}, {Babusiaux}, \&
  {Bailer-Jones}}]{2018arXiv180409365G}
{Gaia Collaboration}, {Brown}, A.~G.~A., {Vallenari}, A., {et~al.} 2018, ArXiv
  e-prints, arXiv:1804.09365

\bibitem[{{Gaia Collaboration} {et~al.}(2016){Gaia Collaboration}, {Prusti},
  {de Bruijne}, {Brown}, {Vallenari}, {Babusiaux}, {Bailer-Jones}, {Bastian},
  {Biermann}, {Evans}, \& et~al.}]{2016A&A...595A...1G}
{Gaia Collaboration}, {Prusti}, T., {de Bruijne}, J.~H.~J., {et~al.} 2016,
  \aap, 595, A1

\bibitem[{{Hollenbach} {et~al.}(1994){Hollenbach}, {Johnstone}, {Lizano}, \&
  {Shu}}]{1994ApJ...428..654H}
{Hollenbach}, D., {Johnstone}, D., {Lizano}, S., \& {Shu}, F. 1994, \apj, 428,
  654

\bibitem[{{Jamialahmadi} {et~al.}(2015{\natexlab{a}}){Jamialahmadi}, {Lopez},
  {Berio}, {Flament}, \& {Spang}}]{2015Ap&SS.355..105J}
{Jamialahmadi}, N., {Lopez}, B., {Berio}, P., {Flament}, S., \& {Spang}, A.
  2015{\natexlab{a}}, \apss, 355, 105

\bibitem[{{Jamialahmadi} {et~al.}(2015{\natexlab{b}}){Jamialahmadi}, {Berio},
  {Meilland}, {Perraut}, {Mourard}, {Lopez}, {Stee}, {Nardetto}, {Pichon},
  {Clausse}, {Spang}, {McAlister}, {ten Brummelaar}, {Sturmann}, {Sturmann},
  {Turner}, {Farrington}, {Vargas}, \& {Scott}}]{2015A&A...579A..81J}
{Jamialahmadi}, N., {Berio}, P., {Meilland}, A., {et~al.} 2015{\natexlab{b}},
  \aap, 579, A81

\bibitem[{{Jamialahmadi} {et~al.}(2018){Jamialahmadi}, {Lopez}, {Berio},
  {Matter}, {Flament}, {Fathivavsari}, {Ratzka}, {Sitko}, {Spang}, \&
  {Russell}}]{2018MNRAS.473.3147J}
{Jamialahmadi}, N., {Lopez}, B., {Berio}, P., {et~al.} 2018, \mnras, 473, 3147

\bibitem[{{Leinert} {et~al.}(2003){Leinert}, {Graser}, {Richichi},
  {Sch{\"o}ller}, {Waters}, {Perrin}, {Jaffe}, {Lopez}, {Glazenborg-Kluttig},
  {Przygodda}, {Morel}, {Biereichel}, {Haddad}, {Housen}, \&
  {Wallander}}]{2003Msngr.112...13L}
{Leinert}, C., {Graser}, U., {Richichi}, A., {et~al.} 2003, The Messenger, 112,
  13

\bibitem[{{Lopez} {et~al.}(2006){Lopez}, {Wolf}, {Lagarde}, {Abraham},
  {Antonelli}, {Augereau}, {Beckman}, {Behrend}, {Berruyer}, {Bresson},
  {Chesneau}, {Clausse}, {Connot}, {Demyk}, {Danchi}, {Dugu{\'e}}, {Flament},
  {Glazenborg}, {Graser}, {Henning}, {Hofmann}, {Heininger}, {Hugues}, {Jaffe},
  {Jankov}, {Kraus}, {Laun}, {Leinert}, {Linz}, {Mathias}, {Meisenheimer},
  {Matter}, {Menut}, {Millour}, {Neumann}, {Nussbaum}, {Niedzielski},
  {Mosonic}, {Petrov}, {Ratzka}, {Robbe-Dubois}, {Roussel}, {Schertl},
  {Schmider}, {Stecklum}, {Thiebaut}, {Vakili}, {Wagner}, {Waters}, \&
  {Weigelt}}]{2006SPIE.6268E..0ZL}
{Lopez}, B., {Wolf}, S., {Lagarde}, S., {et~al.} 2006, in \procspie, Vol. 6268,
  Society of Photo-Optical Instrumentation Engineers (SPIE) Conference Series,
  62680Z

\bibitem[{{Lynden-Bell} \& {Pringle}(1974)}]{1974MNRAS.168..603L}
{Lynden-Bell}, D., \& {Pringle}, J.~E. 1974, \mnras, 168, 603

\bibitem[{{Mulders} {et~al.}(2011){Mulders}, {Waters}, {Dominik}, {Sturm},
  {Bouwman}, {Min}, {Verhoeff}, {Acke}, {Augereau}, {Evans}, {Henning},
  {Meeus}, \& {Olofsson}}]{2011A&A...531A..93M}
{Mulders}, G.~D., {Waters}, L.~B.~F.~M., {Dominik}, C., {et~al.} 2011, \aap,
  531, A93

\bibitem[{{Pani{\'c}} {et~al.}(2014){Pani{\'c}}, {Ratzka}, {Mulders},
  {Dominik}, {van Boekel}, {Henning}, {Jaffe}, \& {Min}}]{2014A&A...562A.101P}
{Pani{\'c}}, O., {Ratzka}, T., {Mulders}, G.~D., {et~al.} 2014, \aap, 562, A101

\bibitem[{{Pineda} {et~al.}(2014){Pineda}, {Quanz}, {Meru}, {Mulders}, {Meyer},
  {Pani{\'c}}, \& {Avenhaus}}]{2014ApJ...788L..34P}
{Pineda}, J.~E., {Quanz}, S.~P., {Meru}, F., {et~al.} 2014, \apjl, 788, L34

\bibitem[{{Quanz} {et~al.}(2015){Quanz}, {Amara}, {Meyer}, {Girard},
  {Kenworthy}, \& {Kasper}}]{2015ApJ...807...64Q}
{Quanz}, S.~P., {Amara}, A., {Meyer}, M.~R., {et~al.} 2015, \apj, 807, 64

\bibitem[{{Quanz} {et~al.}(2013){Quanz}, {Amara}, {Meyer}, {Kenworthy},
  {Kasper}, \& {Girard}}]{2013ApJ...766L...1Q}
---. 2013, \apjl, 766, L1

\bibitem[{{Quanz} {et~al.}(2011){Quanz}, {Schmid}, {Geissler}, {Meyer},
  {Henning}, {Brandner}, \& {Wolf}}]{2011ApJ...738...23Q}
{Quanz}, S.~P., {Schmid}, H.~M., {Geissler}, K., {et~al.} 2011, \apj, 738, 23

\bibitem[{{Ratzka} {et~al.}(2009){Ratzka}, {Schegerer}, {Leinert},
  {{\'A}brah{\'a}m}, {Henning}, {Herbst}, {K{\"o}hler}, {Wolf}, \&
  {Zinnecker}}]{2009A&A...502..623R}
{Ratzka}, T., {Schegerer}, A.~A., {Leinert}, C., {et~al.} 2009, \aap, 502, 623

\bibitem[{{Skrutskie} {et~al.}(1989){Skrutskie}, {Forrest}, \&
  {Shure}}]{1989AJ.....98.1409S}
{Skrutskie}, M.~F., {Forrest}, W.~J., \& {Shure}, M. 1989, \aj, 98, 1409

\bibitem[{{Strom} {et~al.}(1989){Strom}, {Strom}, {Edwards}, {Cabrit}, \&
  {Skrutskie}}]{1989AJ.....97.1451S}
{Strom}, K.~M., {Strom}, S.~E., {Edwards}, S., {Cabrit}, S., \& {Skrutskie},
  M.~F. 1989, \aj, 97, 1451

\bibitem[{{Tatulli} {et~al.}(2011){Tatulli}, {Benisty}, {M{\'e}nard},
  {Varni{\`e}re}, {Martin-Za{\"i}di}, {Thi}, {Pinte}, {Massi}, {Weigelt},
  {Hofmann}, \& {Petrov}}]{2011A&A...531A...1T}
{Tatulli}, E., {Benisty}, M., {M{\'e}nard}, F., {et~al.} 2011, \aap, 531, A1

\bibitem[{{Testi} {et~al.}(2014){Testi}, {Birnstiel}, {Ricci}, {Andrews},
  {Blum}, {Carpenter}, {Dominik}, {Isella}, {Natta}, {Williams}, \&
  {Wilner}}]{2014prpl.conf..339T}
{Testi}, L., {Birnstiel}, T., {Ricci}, L., {et~al.} 2014, Protostars and
  Planets VI, 339

\bibitem[{{van Boekel} {et~al.}(2005){van Boekel}, {Min}, {Waters}, {de Koter},
  {Dominik}, {van den Ancker}, \& {Bouwman}}]{2005A&A...437..189V}
{van Boekel}, R., {Min}, M., {Waters}, L.~B.~F.~M., {et~al.} 2005, \aap, 437,
  189

\bibitem[{{Werner} {et~al.}(2004){Werner}, {Roellig}, {Low}, {Rieke}, {Rieke},
  {Hoffmann}, {Young}, {Houck}, {Brandl}, {Fazio}, {Hora}, {Gehrz}, {Helou},
  {Soifer}, {Stauffer}, {Keene}, {Eisenhardt}, {Gallagher}, {Gautier}, {Irace},
  {Lawrence}, {Simmons}, {Van Cleve}, {Jura}, {Wright}, \&
  {Cruikshank}}]{2004ApJS..154....1W}
{Werner}, M.~W., {Roellig}, T.~L., {Low}, F.~J., {et~al.} 2004, \apjs, 154, 1

\bibitem[{{Wright} {et~al.}(2015){Wright}, {Maddison}, {Wilner}, {Burton},
  {Lommen}, {van Dishoeck}, {Pinilla}, {Bourke}, {Menard}, \&
  {Walsh}}]{2015MNRAS.453..414W}
{Wright}, C.~M., {Maddison}, S.~T., {Wilner}, D.~J., {et~al.} 2015, \mnras,
  453, 414

\end{thebibliography}

\end{document}